\documentclass[%
reprint,
superscriptaddress,
groupedaddress,
bibnotes,
 amsmath,amssymb,
 aps,
prc,
]{revtex4-1}
\usepackage{float}
\usepackage{graphicx}
\usepackage{dcolumn}
\usepackage{bm}

\usepackage{subfigure}
\usepackage{bm, color}
\usepackage{enumitem}
\usepackage{lipsum}
\usepackage[normalem]{ulem}
\usepackage{bbold}

\definecolor{red}{rgb}{0.81,0.13,0.16}

\definecolor{ryangreen}{rgb}{0.20,0.8,0.0}

\usepackage{hyperref}
\hypersetup{
    colorlinks=true,
    linkcolor=blue,
    filecolor=magenta,      
    urlcolor=blue,
    citecolor=blue,
    }

\begin{document}

\preprint{APS/123-QED}

\title{Conformal prediction for uncertainties in nucleon-nucleon scattering}

\author{Habib Yousefi Dezdarani}
\affiliation{Department of Physics, University of Guelph, Guelph, Ontario N1G 2W1, Canada}
\author{Ryan Curry}
\affiliation{Department of Physics, University of Guelph, Guelph, Ontario N1G 2W1, Canada}
\author{Alexandros Gezerlis}
\affiliation{Department of Physics, University of Guelph, Guelph, Ontario N1G 2W1, Canada}

\date{\today}

\begin{abstract}
Conformal prediction is a distribution-free and model-agnostic uncertainty-quantification method that provides finite-sample prediction intervals with guaranteed coverage. In this work, for the first time, we apply conformal-prediction to generate uncertainty bands for physical observables in nuclear physics, such as the total cross section and nucleon-nucleon phase shifts. We demonstrate the method's flexibility by considering three scenarios: (i) a pointwise model, where expansion coefficients in chiral effective field theory are treated as random variables; (ii) a Gaussian-process model for the coefficients; and (iii) phase shifts at various energies and partial waves calculated using local interactions from chiral effective field theory. In each case, conformal-prediction intervals are constructed and validated empirically. Our results show that conformal prediction provides reliable and adaptive uncertainty bands even in the presence of non-Gaussian behavior, such as skewness and heavy tails. These findings highlight conformal prediction as a robust and practical framework for quantifying theoretical uncertainties.
 \end{abstract}

\maketitle

\section{Introduction}
In recent years, nuclear physics has entered a precision era, driven by major advances in \textit{ab initio} many-body methods \cite{Barrett_Navratil_Vary_2013, Carbone_Cipollone_Barbieri_etal_2013, Carlson_Gandolfi_Pederiva_etal_2015, Hagen_Ekstrom_Forssen_etal_2016, Drischler_Holt_Wellenhofer_2021, Hergert_Bogner_Morris_etal_2016, Lynn_Tews_Gandolfi_etal_2019, Lee_2025}, increased computational power, and the wide adoption of chiral effective field theory (EFT) \cite{Epelbaum_Hammer_Meissner_2009, Machleidt_Entem_2011, Gezerlis_Tews_Epelbaum_etal_2013, Gezerlis_Tews_Epelbaum_etal_2014, Epelbaum_Krebs_Meissner_2015, Hammer_Konig_vanKolck_2020}. These developments have significantly expanded the scope of nuclear structure and reaction studies, enabling more accurate and detailed theoretical predictions across a wide range of systems from the lightest nuclei to the neutron matter found in neutron stars \cite{
Hagen_Papenbrock_Hjorth-Jensen_etal_2014, 
Lynn_Tews_Carlson_etal_2016,
Lovato_Gandolfi_Carlson_etal_2016,
Lonardoni_Carlson_Gandolfi_etal_2018, 
Lonardoni_Tews_Gandolfi_etal_2020,
Marino_Jiang_Novario_2024,
Flores_Nollett_Piarulli_2025,
Curry_Somasundaram_Gandolfi_etal_2025, 
Heinz_Miyagi_Stroberg_etal_2025,
Cirigliano_Dekens_Vries_etal_2024,
Gennari_Drissi_Gorchtein_etal_2025}.
As a result, uncertainty quantification (UQ) has become essential to connecting theoretical predictions with experimental data in nuclear physics.  One of the most widely used frameworks for describing low-energy nuclear interactions is chiral EFT, 
which provides powerful and systematic framework for describing interactions based on the symmetries of quantum chromodynamics (QCD).

Efforts have emerged to address these uncertainties \cite{Dobaczewski_Nazarewicz_Reinhard_2014, Ireland_Nazarewicz_2015, Carlsson_Ekstrom_Forssen_etal_2016, Drischler_Melendez_Furnstahl_etal_2020, Duguet_Ekstrom_Furnstahl_etal_2024, Armstrong_Giuliani_Godbey_etal_2025}.  One common approach is to analyze the residual dependence for various cutoffs,  which can provide insight into the size of omitted higher-order terms \cite{Epelbaum_Krebs_Meissner_2015}. 
Recently, a Bayesian framework has been introduced for fitting the
parameters that encode the impact of short-distance physics in chiral EFT, where these parameters are treated as naturally of $\mathcal{O}(1)$ \cite{Schindler_Phillips_2009}. This approach incorporates prior information through Bayes' theorem, the principle of maximum entropy, and employs marginalization over ‘‘nuisance” parameters to construct reliable probability distributions for the parameters of interest.  The Bayesian approach to UQ also helps identify which parts of the calculations (experiment, many-body methods,
EFT Hamiltonians) need improvement to reduce the overall uncertainty \cite{Furnstahl_Phillips_Wesolowski_2015}.  These methods yield degree-of-belief (DoB) intervals and can incorporate prior knowledge and model structure \cite{Wesolowski_Furnstahl_Melendez_etal_2019, Furnstahl_Klco_Phillips_etal_2015, Melendez_Wesolowski_Furnstahl_2017}. Similarly, Gaussian Process (GP) models have been used to treat the expansion coefficients as smooth, correlated functions across inputs such as energy. GPs offer a flexible way to capture correlations and produce credible intervals, while also enabling model diagnostics \cite{Melendez_Furnstahl_Phillips_etal_2019}.
Bayesian methods provide powerful frameworks for uncertainty quantification and rely on assumptions about priors and underlying models.  In this work, we apply conformal prediction (CP) method as a postprocessing step after the Bayesian inference. In this situation, CP does not replace or correct the Bayesian method; instead, it complements it by providing finite-sample prediction intervals with guaranteed coverage for future observations drawn from the same underlying distribution. \cite{Vovk_Gammerman_Shafer_2005, Shafer_Vovk_2008}. Unlike frequentist confidence intervals or Bayesian credible intervals, CP bands provide a finite-sample frequentist guarantee for future data under minimal assumptions. Although CP is a frequentist, model-agnostic method, it can also be applied as a postprocessing step to a Bayesian posterior to construct prediction intervals. The name conformal prediction originates from the use of conformity scores, which assess how well the predicted data points align with the observed data points.
Unlike Bayesian methods, which update prior beliefs based on observed data, CP adopts a fundamentally different philosophy: rather than conditioning on past observations, it aims to quantify uncertainty about future experimental outcomes by constructing prediction intervals that guarantee a specified coverage probability (e.g., 95\% or 68\%) under minimal assumptions. CP provides a powerful tool that does not rely on prior distributions, model correctness, or parametric forms. The standard CP framework assumes that the data are exchangeable, independent, and identically distributed (i.i.d.), which are considerably weaker assumptions than those required by most parametric models. Notably, CP does not require knowledge of the underlying data distribution and remains valid even when the predictive model is not a perfect reflection of the true relationship between $Y$ and $X$, and yet still produces calibrated intervals that contain the true value with a user-specified probability. While conformal prediction is not designed to explicitly model discrepancy, it can inform it by analyzing conformity scores. CP can extend to outlier detection via conformal p values \cite{Bates_Candes_Lei_etal_2023}; a misspecified model fails to capture the underlying data behavior, resulting in systematically large conformity scores. This leads to wide prediction bands, or low conformal p values, which flag data as outliers relative to the model. Thus, CP can provide a fully nonparametric, finite-sample signal of model discrepancy. In contrast, Bayesian approaches such as Bayesian additive regression trees (BART) explicitly model the discrepancy \cite{Chipman_George_McCulloch_2010}.

A key assumption required for this marginal coverage to hold is that our data are exchangeable. Exchangeability means that the joint distribution of data points remains the same under any permutation. In other words, the data are equally likely to appear in any order, drawn from the same distribution. However, when data are not exchangeable, coverage guarantees from standard CP can break down. In such settings, extensions to CP use weighted conformity scores or non-symmetric algorithms to prioritize more recent data points (e.g., by assigning higher weights) \cite{Barber_Candes_Ramdas_etal_2023}.

Furthermore, extensions of the framework can take care of challenges such as distribution shift, model asymmetry, or nonexchangeable data, using tools such as weighted quantiles or residuals. The combination of simplicity, generality, and mathematical robustness makes conformal prediction a powerful and reliable tool for UQ \cite{Lei_Wasserman_2014, Barber_Candes_Ramdas_etal_2023, Angelopoulos_Bates_2023, Bates_Candes_Lei_etal_2023, Burnaev_Vovk_2014, Candes_Lei_Ren_2023, Cauchois_Gupta_Ali_etal_2024, Chernozhukov_Wuthrich_Zhu_2018, Gibbs_Candes_2021, Angelopoulos_Barber_Bates_2025}. In addition, recent studies have explored extending CP to higher dimensions, which is not entirely straightforward \cite{Klein_Bethune_Ndiaye_etal_2025}. While conformal prediction is a powerful statistical tool, it has yet to be widely adopted in physics, though there has been some work in fields such as gravitational-wave searches \cite{Ashton_Colombo_Harry_etal_2024} and in the analysis of astrophysical data \cite{Singer_Williams_Ghosh_2025}.

In this work, we will begin by introducing the theoretical foundation of conformal prediction in Sec.~\ref{Theory}. We will explain how quantiles and quantile regression play a central role in constructing prediction intervals, before explaining how to generate conformal-prediction intervals using toy data. 

In Sec.~\ref{results}, we will first apply the conformal-prediction method to a pointwise model \cite{Melendez_Wesolowski_Furnstahl_2017, Melendez_Furnstahl_Phillips_etal_2019} for nucleon-nucleon scattering cross sections using interactions from chiral EFT, where we analyze the conformity scores and quantile functions to illustrate how CP adapts to varying uncertainty. Next, we will apply our method to samples of nucleon-nucleon cross sections generated using a Gaussian Process framework \cite{Melendez_Furnstahl_Phillips_etal_2019} and show that the CP intervals maintain the desired empirical coverage. 
We then extend our analysis to nucleon-nucleon phase shifts from a realistic local nuclear Hamiltonian \cite{Gezerlis_Tews_Epelbaum_etal_2013, Gezerlis_Tews_Epelbaum_etal_2014, Somasundaram_Lynn_Huth_etal_2024}, across various partial waves and energies to form CP uncertainty bands. Finally, we evaluate the empirical coverage across different energies for specific partial waves and show that our method remains reliable even when the distribution is skewed. 

\section{Theory}\label{Theory}
As noted, conformal prediction provides a statistically rigorous, distribution-free framework for uncertainty quantification. 
When using a machine-learning or statistical model, the predictions made are based on the patterns observed in the training data. However, due to factors such as noisy data, inaccurate models, or limited information, these predictions will always include some inherent uncertainty. One of the key strengths of CP is its model-agnostic nature: it can be applied to a wide range of learning algorithms, including linear regression, or more complex models like Gaussian processes.

\subsection{Quantile regression}
\begin{figure} [t]
	\includegraphics[width=0.5\textwidth]{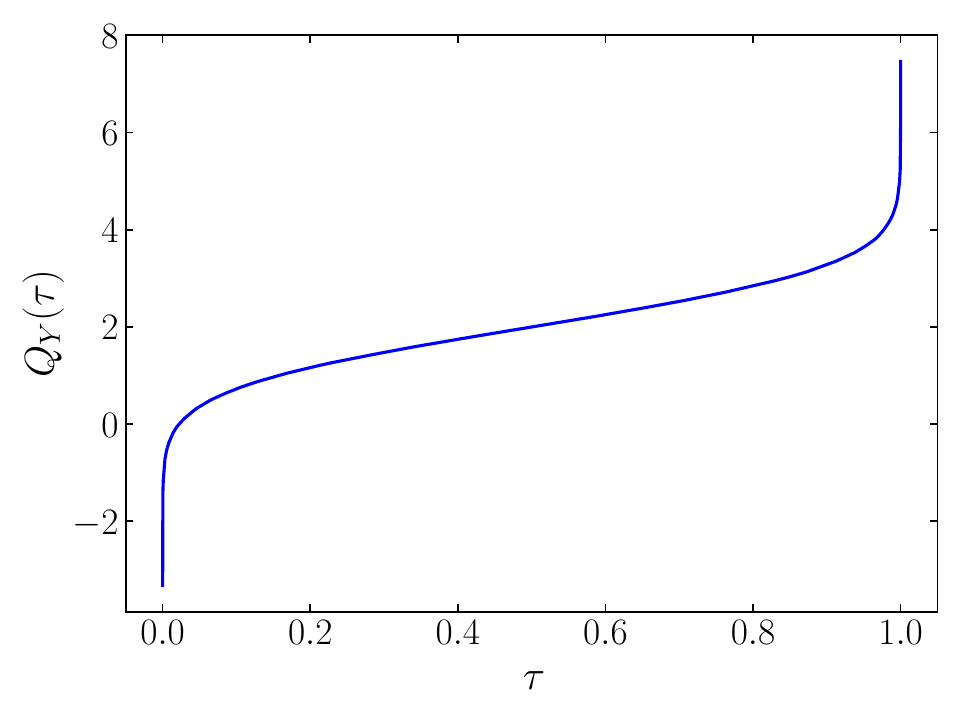} 
	\caption{Empirical quantile function constructed from samples drawn from a normal distribution with mean 2 and standard deviation 1. The curve maps quantile levels $\tau\in[0,1]$ to the corresponding quantiles $Q_Y(\tau)$. This function is the inverse of the cumulative distribution function.}
	\label{fig}
\end{figure}
A key step in the construction of CP intervals is the use of quantiles or quantile regression. Let $Y$ be a random variable with the cumulative distribution function (CDF),
\begin{align}
	F_Y(y) = P(Y)\leq y,
\end{align}
 which gives the probability that the random variable $Y$ is less than or equal to y. The quantile function is then defined as the inverse of the CDF. For a given number $\tau \in[0,1]$, the $\tau$-th quantile is the smallest value $y$ such that the probability of $Y\leq y$ is at least $\tau$ \cite{Davino_Furno_Vistocco_2013}
\begin{align}
	Q_Y(\tau)  = F_{Y}^{-1}(\tau) = \text{inf}\{y:F_Y(y)>\tau\}.
\end{align}

 In Fig.~\ref{fig}, the quantile function identifies the smallest value $y$ such that a certain percentage of data lies below it. For instance, the median corresponds to the $\tau=0.5$ quantile, meaning half of the values fall below it. This approach works well when we study the distribution of a single variable. However, when the distribution of $Y$ depends on another variable $X$, we estimate conditional quantiles, which describe how the quantiles of $Y$ change with some other variable $X$. Quantile regression allows us to estimate these conditional quantiles by modeling the $\tau$-quantile of $Y$ as a linear function of $X$, denoted as $Q_Y(\tau\mid X) = f(X)_{\tau}$.
 
\begin{figure}
	\includegraphics[width=0.5\textwidth]{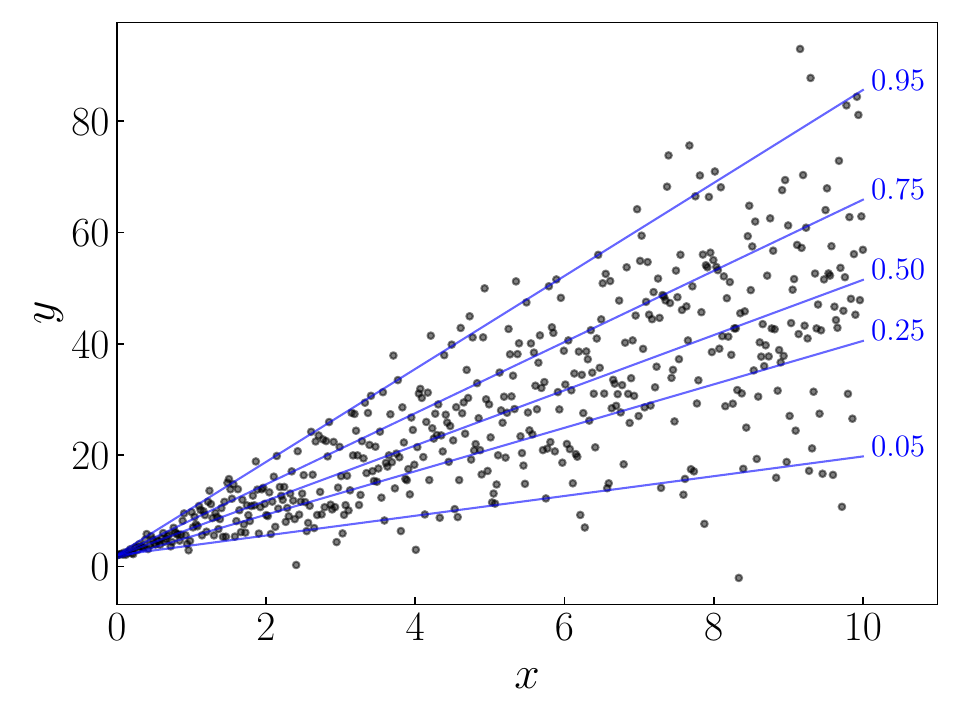} 
	\caption{Quantile regression fits for different quantile levels $\tau\in \{0.05,0.25,0.5,0.75,0.95\}$ applied to synthetic data with heteroscedastic noise. Each fitted line estimates the $\tau$-th conditional quantile $Q_Y(\tau\mid X)$.The model $Y=2 + 3X+X\varepsilon$, where $\varepsilon\sim \mathcal{N}(2,2)$, generates data with increasing variance as $x$ increases. }
	\label{qr}
\end{figure}
To illustrate this procedure, we will first show how quantile regression differs from the simple case of least-squares regression. In least squares, the objective is to estimate some coefficients that minimize the total squared distance between the observed data points and the predicted values:
\begin{align}
	\min{} \sum_{i=1}^n[(y_i - f(x_i)]^2,
\end{align}
which gives the best prediction for the mean of $Y$ given $X$.  In contrast, quantile regression focuses on a specific quantile $\tau$, and minimizes:
\begin{align}\label{quantile-objective}
	\min \sum_{i=1}^n \rho_{\tau}[y_i - f(x_i)],
\end{align}
where $\rho_{\tau}(u)$ is the loss function 
\begin{align}
	\rho_{\tau}(u)=
	\begin{cases}
		\tau u  & \text{if } u \geq 0\\
		(\tau - 1)u & \text{if } u < 0,
	\end{cases}
\end{align}
and $u$ is the residual or difference between the observed values and predicted values. The loss function says that when $\tau=0.5$, positive and negative residuals are weighted the same, which means that we want the same number of data points above and below the line. For $\tau > 0.5$, positive residuals are weighted more, meaning there would be more data points under the line. And for $\tau=0.1$, data points above the line are penalized more compared to those located below the line.

To demonstrate how quantile regression works in practice, we generate synthetic data where the noise is heteroscedastic, meaning the variance of the noise depends on the variable $X$. Specifically, we use the model,
\begin{align}
	Y = 2 + 3X + X  \varepsilon, \hspace{0.4cm} 
\end{align}
where $\varepsilon$ is normally distributed with mean 2 and standard deviation 2, implying a variance of Var($\varepsilon$) = $\sigma^2$ = $4$. Since the noise term is scaled by $X$, the variance of $X \varepsilon$ becomes Var($X \varepsilon$) = $\sigma^2 X^2$ = $4 X^2$. This input-dependent noise, i.e., heteroscedasticity, results in larger variability in $Y$ as $X$ increases.

We show the quantile regression fits in Fig. \ref{qr}, which are obtained by solving the minimization of Eq.~(\ref{quantile-objective}), for various values of $\tau \in \{0.05,0.25,0.5,0.75,0.95\}$. The median regression line ($\tau = 0.5$) divides the data points into two halves, while the lower ($\tau = 0.05$) and upper ($\tau = 0.95$) quantile lines capture the spread of the distribution. For instance, the quantile line corresponding to $\tau = 0.95$ shows that for higher values of $y$, the rate of increase with respect to $x$ is significantly greater than for lower values of $y$. This visualization emphasizes the strength of quantile regression in capturing the behavior of data across different quantiles. In the next section, we will show how to use these quantiles to construct CP intervals.

\subsection{Conformal prediction intervals}
To see the strength of the conformal-prediction approach, suppose we begin with two random variables, $X$ and $Y$, where $X$ represents the independent variable and $Y$ the dependent variable. Their possible values are denoted by $x_i$ and $y_i$ drawn from an unknown distribution, for $i=1,2,...,n$. The goal here is not just to provide a point estimate $f(x_{n+1})$, but to construct a prediction set $C(x_{n+1})$, such that
\begin{align}\label{marginal-coverage}
	\mathcal{P}[y_{n+1}\in C(x_{n+1})]\geq 1- \alpha,
\end{align}
which states that the probability a new $y_{n+1}$ is within the prediction set, generated by $C(x_{n+1})$, is greater than or equal to $ 1-\alpha$. This statement is known as the marginal coverage property \cite{Angelopoulos_Barber_Bates_2025}. The story regarding conditional inference guarantees is more complicated \cite{Barber_Candes_Ramdas_etal_2023}.
 Here $\alpha \in (0, 1)$ is a user-defined error level, and $1-\alpha$ is the desired coverage level. For example, if the user specifies $\alpha=0.05$, the method constructs prediction sets that contain the true $y_{n+1}$ at least 95\% of the time. In the following, we review the formalism introduced by \cite{Angelopoulos_Barber_Bates_2025}.
A key assumption required for this marginal coverage to hold is that our data are exchangeable. Exchangeability means that the joint distribution of data points remains the same under any permutation. In other words, the data are equally likely to appear in any order. In the CP, exchangeability is applied as the fundamental assumption that enables coverage guarantees. It is used when computing conformity scores during calibration, under the assumption that the calibration and test data points are drawn from the same distribution. However, when data are not exchangeable, coverage guarantees from the standard CP can break down. In such settings, extensions to the CP use weighted conformity scores or non-symmetric algorithm to prioritize more recent data points (e.g., by assigning higher weights to recent data points). These techniques make the CP more robust to non-exchangability, allowing the prediction intervals to maintain the coverage guarantee.

In addition, the conformal-prediction method does not require assumptions about the correctness of the predictive model or the data distribution. 
Even if the model is poorly fitted to the data, the resulting prediction sets will be large, representing high uncertainty, whereas a well-fitted model yields smaller prediction intervals. This makes CP a powerful tool to construct reliable and robust prediction intervals. 
\begin{figure}
	\includegraphics[width=0.5\textwidth]{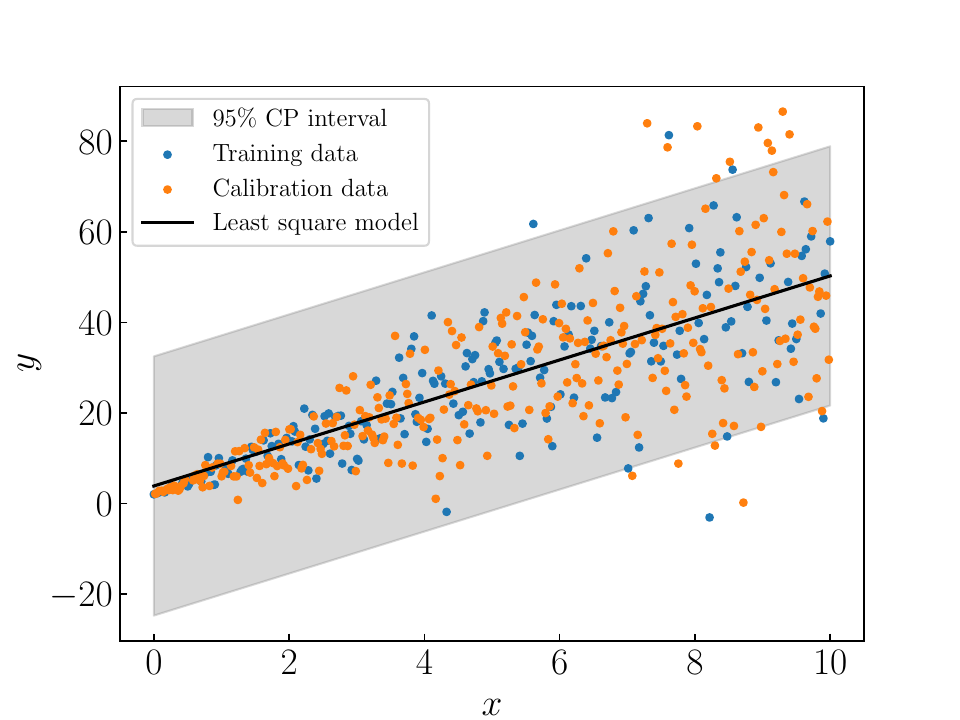} 
	\caption{95\% naive CP interval for a toy dataset including 500 samples with heteroscedastic noise. The prediction model is a least squares regression fitted to the training data, and the shaded region represents the CP interval constructed using absolute residuals from the calibration data. The black line shows the fitted model. The empirical coverage is 0.9454.}
	\label{cp-toy1}
\end{figure}
We start by considering a simple, and somewhat naive, implementation of conformal prediction known as split conformal prediction. In this method, data are divided into two subsets: the training set, with $n_{\text{train}}$ data points, is used to fit a predictive model, and the calibration set, with $n_{\text{calib}}$ data points, is used to evaluate the model’s performance via conformity scores.
Conformity scores measure how well a predicted value conforms to the true observed value. A common conformity score is the absolute residual between the prediction and the observed data:
\begin{align}
    S = |Y - {f}(X)|. 
\end{align}
The naive split conformal prediction approach proceeds as follows:
\begin{enumerate}[label=(\roman*)]
	\item Train a predictive model $f$ using the training data. This model could be any regression model (e.g., linear regression).
	\item Compute conformity scores on the calibration set.
        \item Arrange these scores in ascending order, and define $q$ as the $\left[ (1 - \alpha)(n_{\text{calib}} + 1)\right]$th element of this ordered sequence.
    
	\item Construct the naive prediction interval for a new input $x_{n+1}$ as:
\end{enumerate}
    \begin{align}
	C(x_{n+1}) = [f(x_{n+1}) - q,\ f(x_{n+1}) + q ],
\end{align}
If the data are independent and identically distributed, this method guarantees that the prediction interval will contain the true data $y_{n+1}$, with probability $1-\alpha$, regardless of the underlying distribution.

\begin{figure}
	\includegraphics[width=0.5\textwidth]{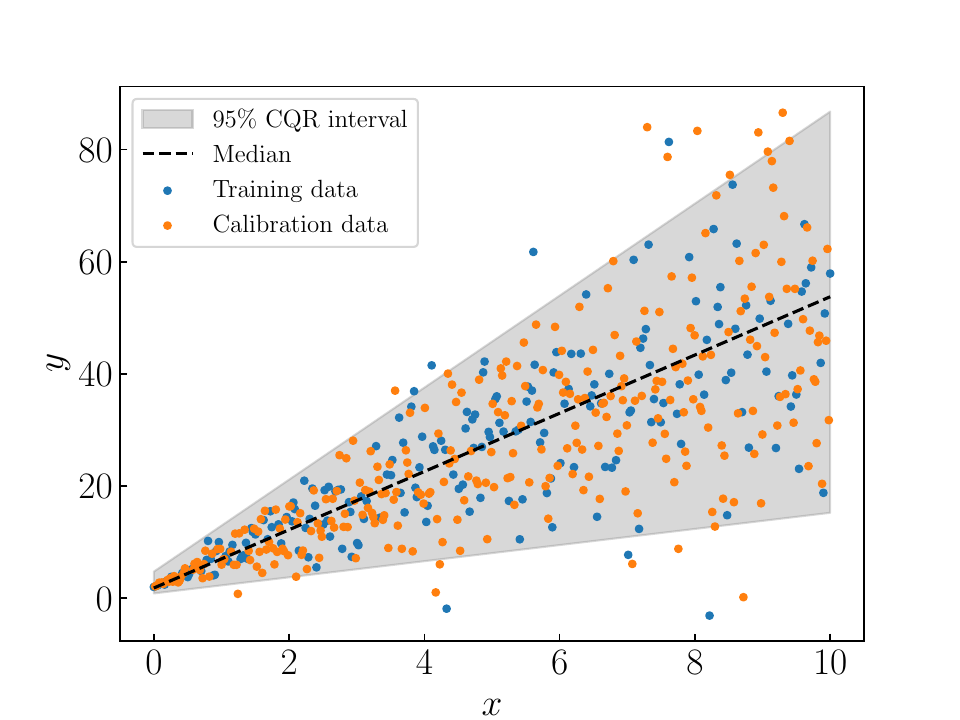} 
	\caption{95\% CQR interval for a toy dataset including 500 samples with heteroscedastic noise. The prediction model is a quantile regression model trained to estimate the conditional quantiles of the data. The shaded region represents the conformalized quantile regression interval constructed by adjusting the initial quantile bounds using conformity scores from the calibration data. This approach produces asymmetric prediction bands that adapt to the variability of the data. The dashed line indicates the median prediction. The empirical coverage is 0.9499. }
	\label{toy-cqr95-band}
\end{figure}
\begin{figure*}[t]
	\includegraphics[width=1\textwidth]{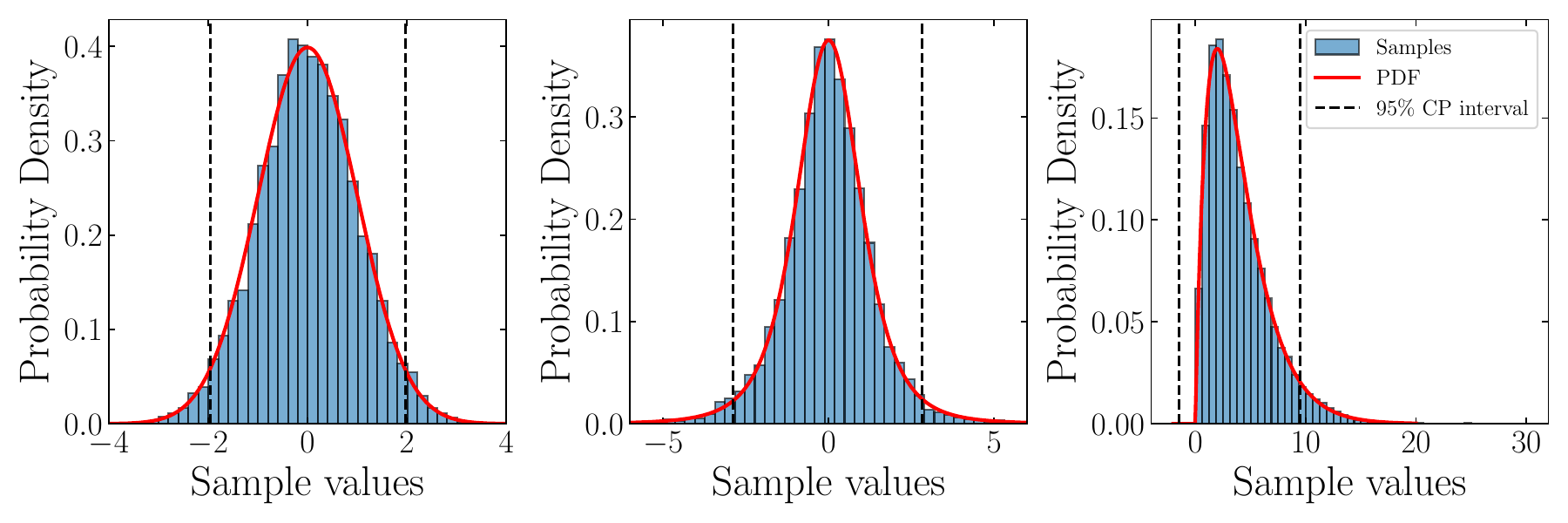} 
	\caption{Comparison of 95\% naive CP intervals for three distributions: (a) Gaussian, (b) student-$t$ with four degrees of freedom, and (c) chi-squared with 4 degrees of freedom. Each subplot shows the histogram of 10000 sample values with the true distribution. The empirical coverage for each plot is calculated 0.9495, 0.9496, and 0.9499, respectively.}
	\label{cp-toy}
\end{figure*}
In Fig. \ref{cp-toy1} we plot the naive 95\% conformal-prediction interval constructed using the split conformal-prediction method applied to a toy dataset with heteroscedastic noise. Here, a simple linear model is trained on the training data, and the prediction interval is constructed based on the absolute residuals using a calibration set. As we can see, the resulting prediction band is symmetric around the model prediction and has a constant width across all $x$ values. This is because our choice for the conformity score, $S = |Y - {f(X)}|$, does not account for variations in the noise. 
While the absolute residual conformity score provides a simple and intuitive way to construct conformal prediction intervals, it always produces symmetric intervals, which may not be ideal when the data distribution is skewed or the variance is not constant. To address this, a more flexible approach is to use Conformalized Quantile Regression (CQR) which allows for asymmetric intervals. 

The key idea behind CQR is to first use quantile regression to estimate the lower and upper quantiles of $Y$ given a value $X$. For a given value $\alpha$, the lower and upper quantiles are defined as :
\begin{align}
Q_Y(\alpha/2\mid X)\ \text{and}\ Q_Y(1- \alpha/2\mid X)
\end{align}
 These quantiles define an initial prediction interval. However, since these estimates are not always reliable, we apply a conformal correction. For a calibration set, we compute conformity scores:
\begin{align}
	S = \text{max} \{Q_Y(\alpha/2\mid X)-Y,\ Y - Q_Y(1- \alpha/2\mid X) \}, 
\end{align}
which measures how far the true value $Y$ lies outside the predicted quantile interval at level $\alpha$. After computing $q$, using step 3 of the split conformal prediction procedure outlined above, finally define the CQR prediction interval for a new $x_{n+1}$ as:
\begin{align}
C(x_{n+1}) = 
[ Q_y\left(\tfrac{\alpha}{2} \mid x_{n+1} \right) - q,\ 
      Q_y\left(1 - \tfrac{\alpha}{2} \mid x_{n+1} \right) + q ],
\end{align}

\begin{figure} [b]
	\centering
	\includegraphics[width=0.5\textwidth]{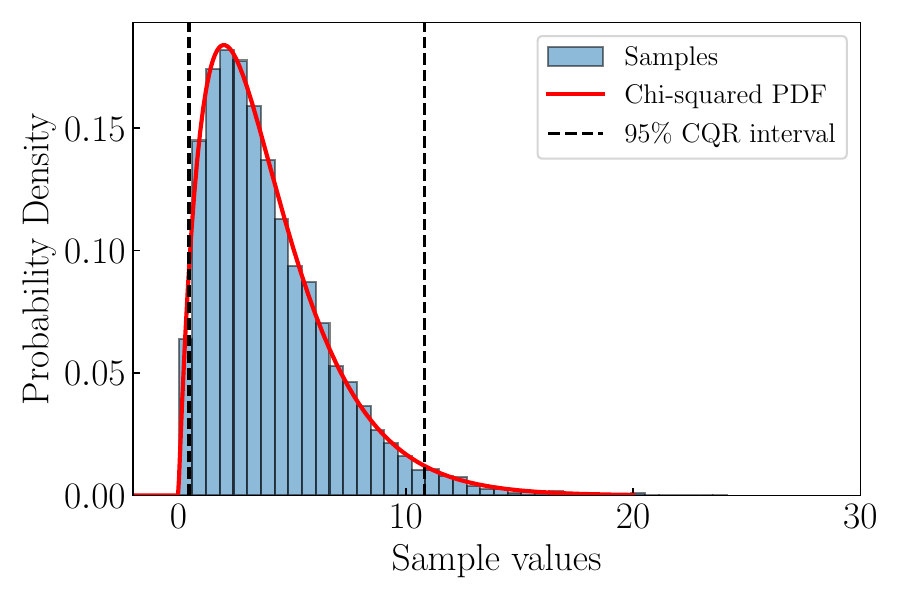} 
	\caption{95\% CQR interval for toy data drawn from a chi-squared distribution with 4 degrees of freedom. Unlike the naive CP method, CQR adapts this interval, ensuring meaningful prediction bands.  }
	\label{toy-samples-cqr}
\end{figure}
where the training data points $n_{\text{train}}$ are used to estimate the lower and upper quantiles. This conformal-prediction method adjusts the initial quantile interval by expanding it when $q$ is positive, or shrinking when $q$ is negative. The 95\% prediction band created using the CQR method in shown in Fig. \ref{toy-cqr95-band}. Compared to the standard residual-based conformal prediction, CQR produces asymmetric bands that adjust to the variability of the data. We can see that the band becomes wider as $x$ increases, which reflects the increasing noise.

Suppose we are given a set of posterior samples for a physical observable generated at a specific energy. Our goal is to construct a prediction interval that captures the true value with a desired coverage. To simulate this realistic setup, we use toy data sets drawn from three distinct distributions such as Gaussian, Student-t, and chi-squared, then apply the conformal procedure to each. The prediction model is defined as the mean of the training data points:
\begin{align}
	\bar{y} = \frac{1}{n} \sum_{i=1}^{n}y_i.
\end{align}
To build naive prediction intervals using absolute residual conformity scores, we follow the steps outlined above. Then the naive prediction intervals for any new point can be calculated as:
\begin{align}
	C = [\bar{y} - q,\bar{y} + q].
\end{align}

In Fig. \ref{cp-toy} we compare the performance of the naive CP method under these distinct data distributions. For each case, the histogram represents the sample values, and the red line shows the true probability distribution function of each. The black vertical dashed lines show the constructed 95\% prediction interval. Despite the difference in the shape of the distributions, notably the skewness in the chi-squared case, the conformal intervals still provide reliable coverage. This highlights one of CP's most important features: it makes no assumptions about the underlying distribution, and yet provides robust uncertainty bands. The only problem is that, when we apply the naive CP to chi-squared samples, the resulting prediction intervals can sometimes extend into negative values, which are not reasonable. To address this issue, we again turn to CQR, which adapts the prediction intervals to better respect the distribution's shape. As shown in Fig. \ref{toy-samples-cqr}, using CQR ensures that the prediction intervals for the chi-squared case remain non-negative, providing a meaningful uncertainty band. In the following applications, we repeat the CQR procedure at each energy value, using the posterior samples to construct CP intervals.

\section{Applications}\label{results}
\begin{figure} 
\includegraphics[width=0.5\textwidth]{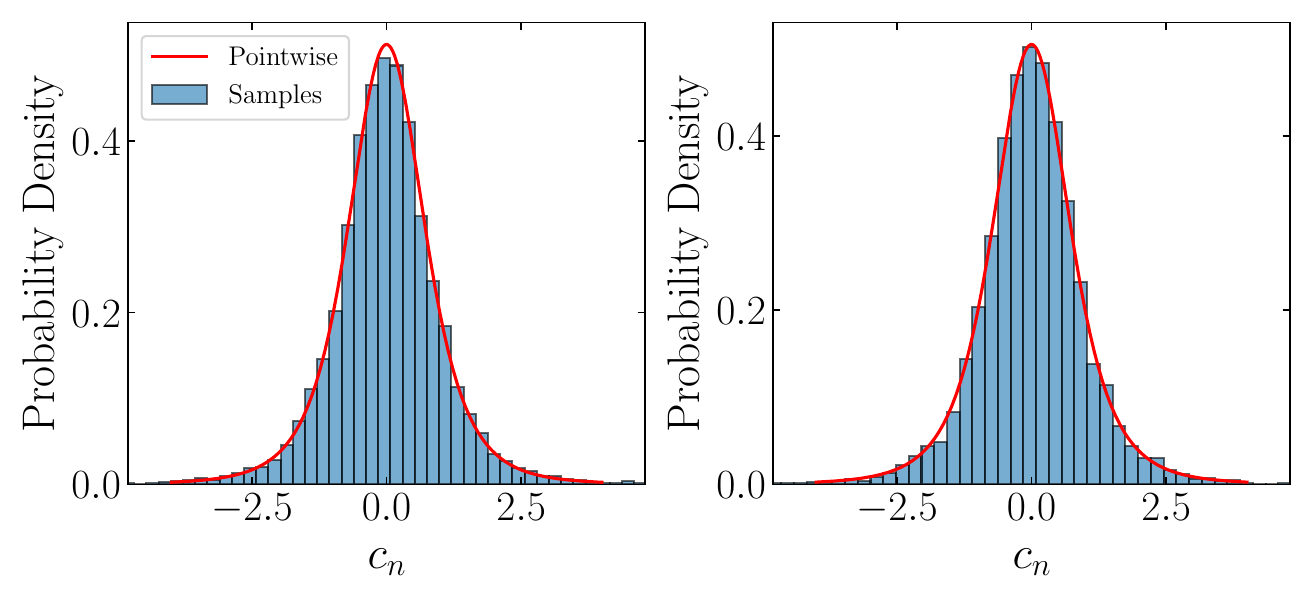} 
\caption{Posterior distributions of the expansion coefficient $c_n$ at the N$^2$LO order, obtained from the pointwise model, for the total cross section at two energies (50 and 200 MeV), using the EKM potential for $R_0= 0.9\ \text{fm}$. The distributions are created using 10000 samples.}
\label{coeffs-pointwise}
\end{figure}
To quantify uncertainties in nucleon-nucleon scattering observables such as the total cross section, we apply conformal prediction as a postprocessing step to construct uncertainty bands with guaranteed coverage. In order to test our approach, we start by considering posterior distributions of the total cross section, obtained either from the pointwise model or the Gaussian-process (GP) model, both due to BUQEYE Collaboration \cite{Melendez_Furnstahl_Phillips_etal_2019}. We apply the CQR approach to these posteriors to obtain distribution-free prediction intervals. 

We then compute nucleon-nucleon phase shifts for realistic local nuclear Hamiltonians that are commonly used in many-body calculations such as quantum Monte Carlo. We are able to use CP to generate uncertainty bands with guaranteed coverage across all partial waves and laboratory energies. We also discuss how we calculate the phase shifts as well as how our CQR approach, which guarantees empirical coverage, could be used in the future when fitting these interactions. 

\subsection{Pointwise model}
The pointwise model, as introduced by the BUQEYE Collaboration \cite{Melendez_Furnstahl_Phillips_etal_2019}, provides a novel method for quantifying truncation errors in chiral EFT observables. Here, we apply their ideas and terminology to generate posterior distributions of nucleon-nucleon cross sections. In chiral EFT, physical observables are expressed as an expansion in powers of $Q$, which is $y = y_{\text{ref}} \sum_{n=0}^{\infty}c_n Q^n.$
\begin{figure} 
\includegraphics[width=0.5\textwidth]{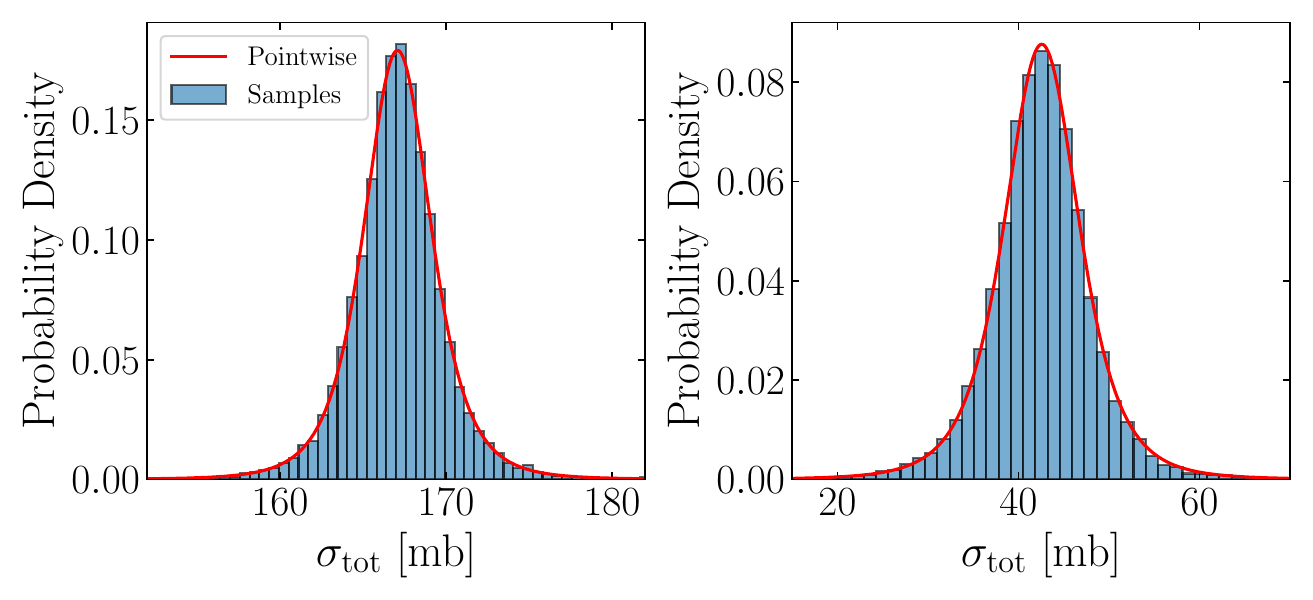} 
\caption{Posterior distributions for the total cross section at N$^2$LO, obtained from the pointwise model using EKM potential for $R_0= 0.9\ \text{fm}$ at energies 50 and 200 MeV. The histograms are created using 10000 posterior samples.}
\label{sgt N2LO-dist-pointwise}
\end{figure}
Here $Q$ is the ratio of the typical momenta of the nucleons over the momenta at which the theory is expected to breakdown. $y_{\text{ref}}$ is the natural size of $y$ and generally could be taken to be the leading-order estimate $y_0$. The $c_n$ parameters are dimensionless expansion coefficients expected to be of natural size.
As a first step, we assume that an unseen expansion coefficient $c_n$ is drawn independently from a Student-t distribution with degrees of freedom $\nu$, mean 0, and standard deviation $\tau$,

\begin{align}
c_n \mid \bm {{c}_k} \sim t_\nu(0, \tau^2).
\end{align}
The posterior parameters $\nu$ and $\tau$ in the Student-t distribution are updated based on the known coefficients $\bm {{c}_k}$. They are computed as:

\begin{align}
    \nu = \nu_0 + n_c,\hspace{1cm} \nu\tau^2 = \nu_0\tau_0^2 + \bm {{c}_k}^{~2}
\end{align}
where $\nu_0$ and $\tau_0$ are prior hyperparameters, and $n_c$ is the number of known coefficients in $\bm {{c}_k}$. Here, we use $\nu_0=1$ and $\tau_0=1$, and update  $\nu$ and $\tau$ before sampling from the posterior \cite{Melendez_Furnstahl_Phillips_etal_2019}. 

\begin{figure*}[t]
    \centering
    \includegraphics[width=0.45\textwidth]{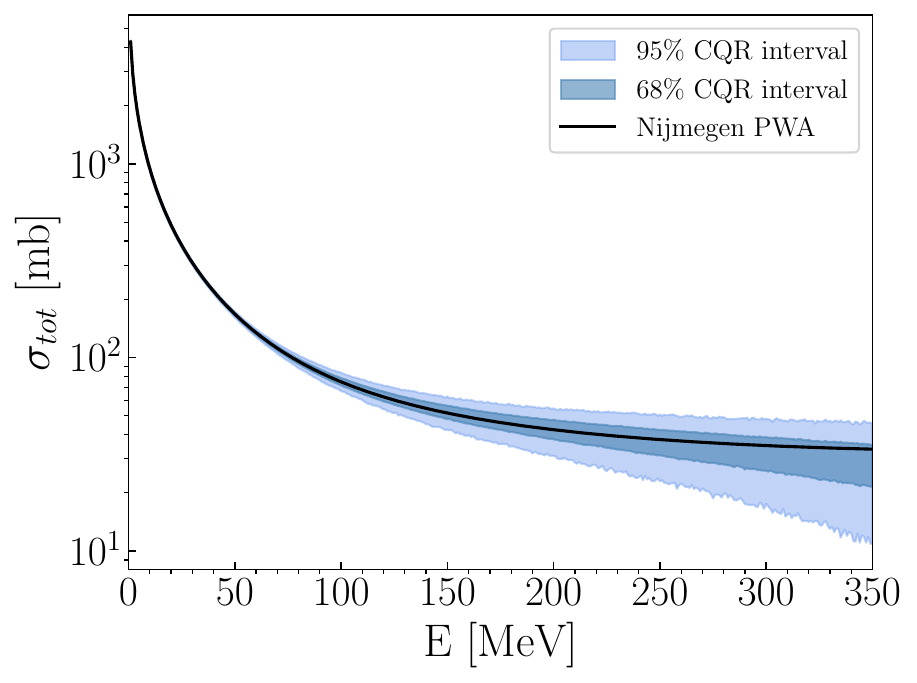}
    \hspace{0.05\textwidth}
    \includegraphics[width=0.45\textwidth]{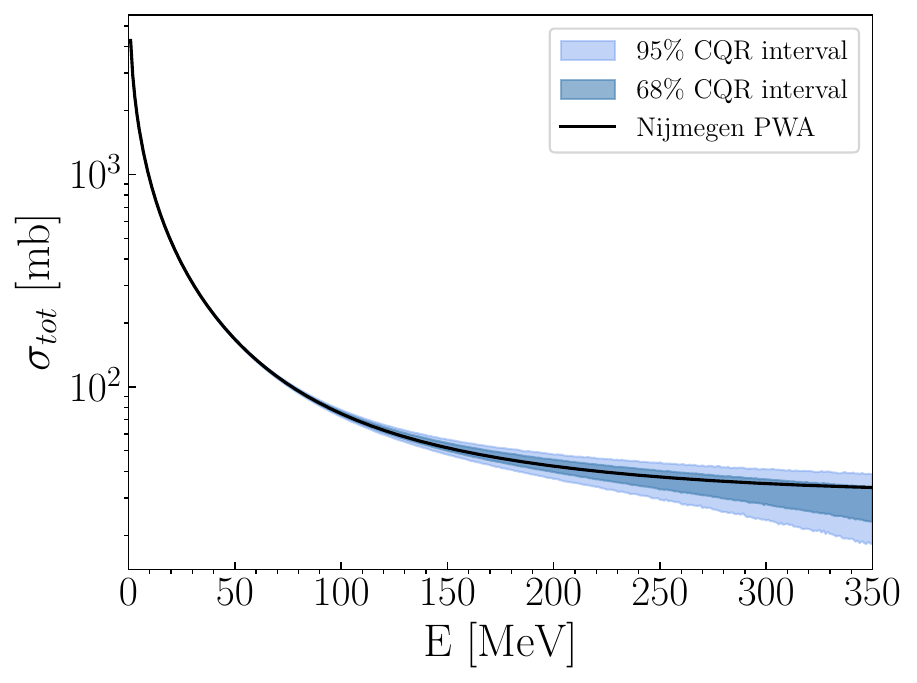}
    \caption{CQR intervals obtained for the total cross section $\sigma_{\text{tot}}$ at N$^2$LO using the EKM potential at $R_0=0.9\,\text{fm}$, shown across a range of energies. The left panel corresponds to the pointwise model, while the right panel corresponds to the Gaussian-process (GP) model. Dark and light blue bands denote the 68\% and 95\% CQR intervals constructed from 10000 samples, respectively. In both cases, the width of the intervals increases with energy. As we can see, the bands are jagged because of fluctuations in the underlying samples; this effect is stronger for the pointwise model, where the samples are noisier, and weaker for the GP, where the correlations across energies smooth behavior.}
    \label{sgt-cp-intervals}
\end{figure*}
\begin{figure*}
    \centering
    \includegraphics[width=0.45\textwidth]{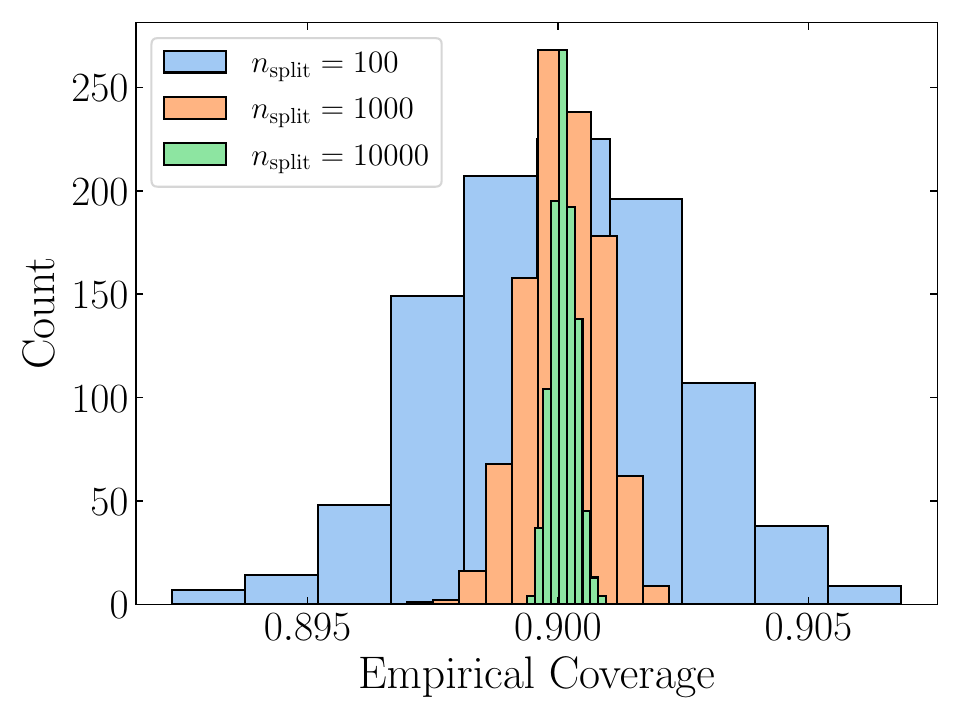}
    \hspace{0.05\textwidth}
    \includegraphics[width=0.45\textwidth]{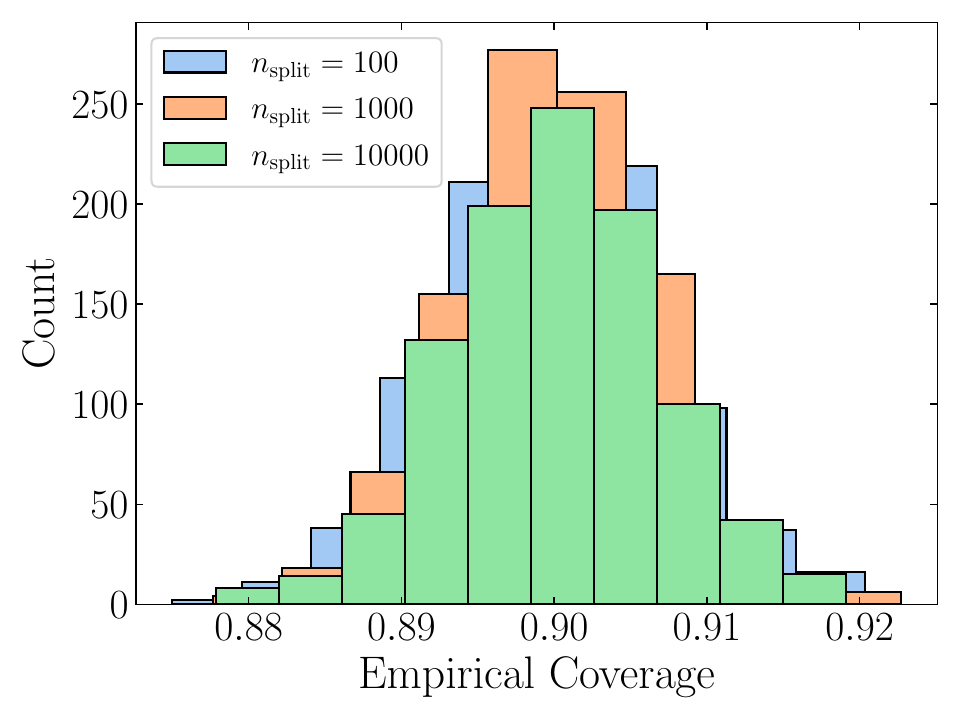}
    \caption{Empirical coverage distribution for the total cross section $\sigma_{\text{tot}}$ at N$^2$LO and $E=50$ MeV using the EKM potential at $R_0=0.9\,\text{fm}$ for three different values of  ($n_{\text{split}}=100,1000,10000$). The left panel shows the empirical coverage distribution for CQR using 2000 samples calculated from the GP, with $ n_{\text{calib}}=900$,  $n_{\text{train}}=900$, and $n_{\text{valid}}=200$. For each $n_{\text{split}}$, a new CQR interval is constructed, and this process is repeated $n_{\text{trial}}=1000$ times (where each trial uses freshly drawn samples). As $n_{\text{split}}$ increases, the spread of empirical coverage narrows and converges toward the target coverage $0.9$. The right plot illustrates the empirical distribution for  DoB intervals, which remain fixed across different $n_{\text{split}}$, as they are not based on the calibration and training data points.}
    \label{empirical}
\end{figure*}

To study the behavior of the expansion coefficient $c_n$, we draw samples from the Student-t posterior for each energy. Here, we take the data for nucleon-nucleon scattering, which are available on 's GitHub repository \cite{buqeye-repo}. In Fig. \ref{coeffs-pointwise} we plot the resulting posterior distributions for $c_n$ at two energies (50 and 200 MeV), based on the EKM potential \cite{Epelbaum_Krebs_Meissner_2015} with a cutoff of $R_0= 0.9\ \text{fm}$. The blue histograms show the samples drawn from the pointwise posterior, and the red line represents the corresponding analytic Student-t probability density function. In terms of the EFT observables, the posterior distribution for the full prediction $ y =y_k + \delta y_k$ is \cite{Melendez_Furnstahl_Phillips_etal_2019},

\begin{align}
y \mid \bm {{y}_k}, Q \sim t_\nu \left( y_k,\ y_{\text{ref}}^2 \frac{Q^{2(k+1)}}{1 - Q^2}  \tau^2 \right)
\end{align}
Here, Q is again the expansion parameter and $y_{\text{ref}}$ is the reference scale. Next, we generate samples for the full observable $y$ at each energy using the posterior defined above. In Fig. \ref{sgt N2LO-dist-pointwise} we show the distributions of the total cross section at the mentioned energies using EKM and for $R_0= 0.9\ \text{fm}$. The histogram corresponds to the empirical density of the samples and red curves indicate the analytic Student-t PDF. At 50 MeV, the distributions are narrow and roughly symmetric, centered around the predicted value $y_k$. This suggests that the uncertainty is small at low energies, where the expansion parameter Q is small. As we go to higher energies, especially at 200 MeV, the distribution is clearly broader. This indicates that the uncertainty becomes larger at higher energies. 

\begin{figure}
\includegraphics[width=0.49\textwidth]{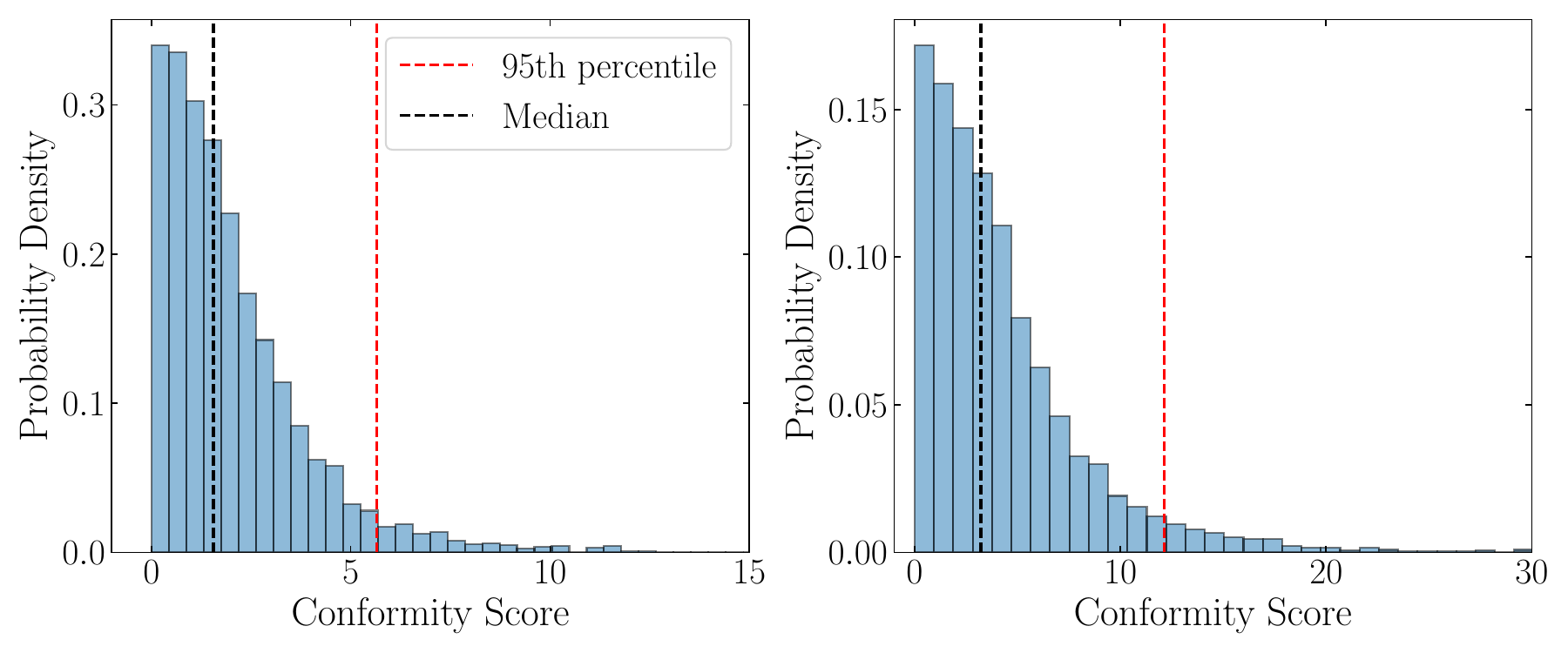} 
\caption{Histogram of the absolute residual conformity scores obtained from the pointwise model for the calibration set at two energies (50 and 200 MeV). The red dashed lines indicate the 95th percentile, and the black dashed line shows the median of the conformity scores.}
\label{confr-N2LO,pointwise}
\end{figure}

\begin{figure}[b]
\includegraphics[width=8cm, height=6.2cm]{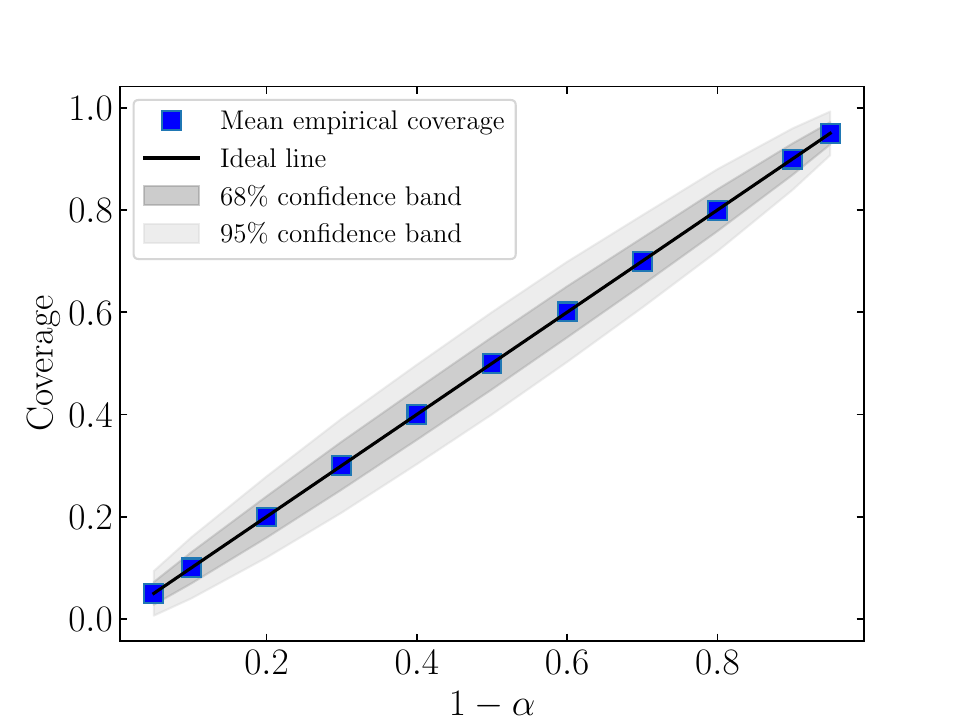} 
\caption{Empirical coverage based on the pointwise model for conformal prediction intervals for $n_{\text{split}}=4000$ at $E=50\ \text{MeV}$ as a function of $1-\alpha$. Blue squares show the mean of empirical coverage, and the light and shaded bands represent the standard deviation 1$\sigma$ and $1.96 \ \sigma$. }
\label{coverage-N2LO-Pointwise}
\end{figure} 

In Fig. \ref{sgt-cp-intervals} we show the CQR intervals for the total cross section $\sigma_{\text{tot}}$ at N$^2$LO order using the EKM potential with $R_0= 0.9\ \text{fm}$ across a wide range of energies (we will discuss the right panel of Fig. \ref{sgt-cp-intervals} in the following section). The dark and light blue bands represent the 68\% and 95\% CQR intervals, respectively, and the black line represents the experimental results from the Nijmegen partial wave analysis \cite{Stoks_Klomp_Rentmeester_etal_1993}. As the energy increases, the intervals become wider, reflecting the increasing uncertainty in the EFT predictions. At low energies, the cross sections are larger and the prediction intervals are relatively narrower. In contrast, at higher energies, the intervals broaden significantly, capturing the increased uncertainty. This shows how the CQR method adaptively accounts for model uncertainty without relying on specific assumptions about the underlying distribution.

In Fig. \ref{empirical} we show a side-by-side comparison of empirical coverage distributions for the total cross section $\sigma_{\text{tot}}$ at N$^2$LO and $E=50$ MeV for $n_{\text{split}}=100,1000,10000$, where $n_{\text{split}}$ refers to the number of times the dataset is randomly split into training, calibration, and validation sets.  In the left panel we show the empirical coverage distributions for CQR. A dataset of 2000 samples was generated using a GP and divided into $n_{\text{train}}=900$, $n_{\text{calib}}=900$, and $n_{\text{valid}}=200$. For each value of $n_{\text{split}}$, a new CQR interval was formed, and the entire procedure was repeated $n_{\text{trial}}=1000$ times. Here $n_{\text{trial}}$ denotes the number of independent trials, each carried out with freshly drawn data from the GP.  The results show that increasing $n_{\text{split}}$ reduces the spread of the empirical coverage distribution, which becomes narrower and centered around the target coverage $0.9$. In the right panel of Fig. \ref{empirical} we show the empirical coverage distribution of the DoB under the same setup. Unlike the CQR results in the left panel, the DoB interval does not narrow with increasing $n_{\text{split}}$, since the DoB interval is fixed and does not depend on calibration data points. As a result, the three distributions overlap rather than getting narrower. This result demonstrates the validity guarantee promised by conformal prediction: as the number of times we have randomly split the data ($n_{\text{split}}$) increases, the empirical coverage distribution narrows toward the target coverage.

In Fig. \ref{confr-N2LO,pointwise} we explore how the absolute residual conformity scores are distributed at different energies. The black dashed line represents the median, and the red dashed line shows the 95th percentile of the conformity scores, denoted as $q$, which defines the width of the conformal prediction interval. As we can see, from 50 MeV up to 200 MeV, the tail of the distribution becomes longer and the value of $q$ increases. This reflects greater uncertainty in the data at higher energies. As a result, the prediction intervals become wider at higher energies.

To evaluate the validity of the conformal prediction intervals, we plot the empirical coverage to the target coverage levels across a range of $1-\alpha$. The results of this evaluation for the total cross section at $E=50\ \text{MeV}$ based on $n_{\text{split}}= 4000$, are shown in Fig. \ref{coverage-N2LO-Pointwise}.
The empirical coverage values closely follow the ideal line, always remaining within the 68\% and 95\% confidence band. These results validate the robustness and reliability of our method and show that conformal prediction provides guaranteed coverage with minimal assumptions.

\subsection{Gaussian Processes}
A Gaussian process (GP) is a flexible, nonparametric model that can be used in regression problems. Here, we adopt the  strategy \cite{Melendez_Furnstahl_Phillips_etal_2019}, where  the expansion coefficients $c_n(x)$ are treated as random functions that are drawn from a GP distribution:
\begin{align}
	c_n(x) \sim \mathcal{GP}[0,\overline{c}^2r(x,x';l) ],
\end{align}
where $\overline{c}^2$ stands for the marginal variance that controls the overall size of the coefficients, and $r(x,x';l)$ represents the kernel function or correlation function, which is chosen to be a radial basis function (RBF) kernel defined as follows,
\begin{align}
	r(x,x';l) = \text{exp}\left(-\frac{(x-x')^2}{2l^2}\right).
\end{align}
Here, $l$ is the length scale, which controls how quickly the function can change. In addition, a scaled inverse-chi-squared prior is considered for the $\overline{c}^2$, which is a conjugate prior for the variance of the GP. For more details about choosing the priors and hyperparameters in GP, see Refs. \cite{Furnstahl_Klco_Phillips_etal_2015, Melendez_Wesolowski_Furnstahl_2017, Melendez_Furnstahl_Phillips_etal_2019}.
\begin{figure}[t]
\includegraphics[width=0.49\textwidth]{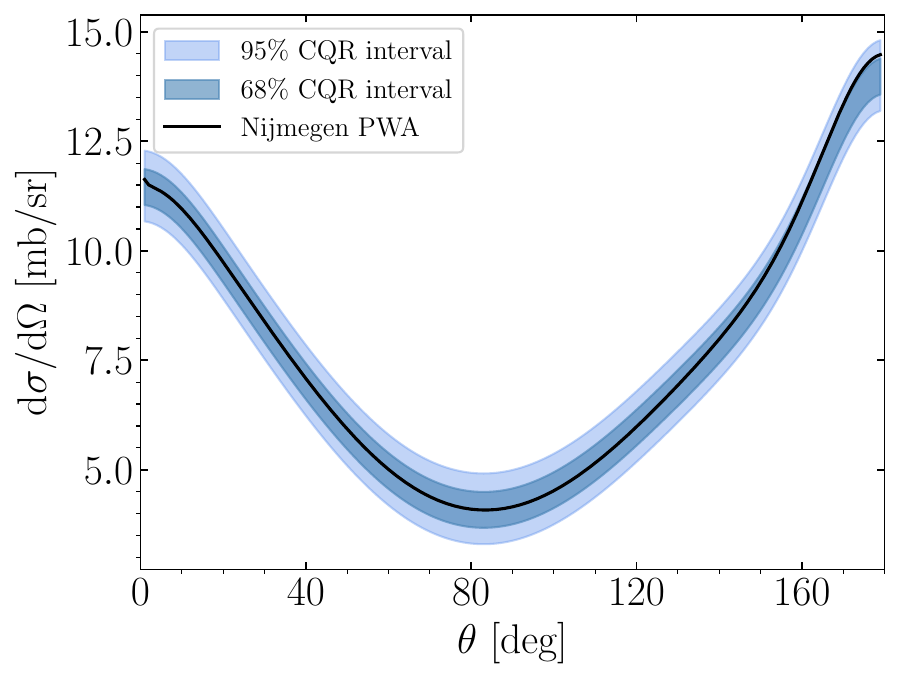} 
\caption{CQR intervals for the differential cross section as a function of polar scattering angle, computed at N$^2$LO and $E=96\ \text{MeV}$ using EKM potential with $R_0= 0.9\ \text{fm}$. The dark and light blue bands represent the $68\%$ and $95\%$ CQR intervals, respectively. The black curve corresponds to the Nijmegen PWA result \cite{Stoks_Klomp_Rentmeester_etal_1993}. The prediction bands are obtained by sampling the chiral EFT expansion coefficients using the GP model and applying CQR to construct distribution-free prediction intervals constructed from 10000 samples.}
\label{dsg-CP-bands}
\end{figure}

After sampling the total cross section expansion coefficients at N$^2$LO from the GP model across different ranges of energies, we draw samples for the total cross section, $\sigma_{\text{tot}}$, at each energy value. Using these samples, we then apply CQR to construct a distribution-free prediction interval. In the right panel of Fig. \ref{sgt-cp-intervals} the resulting 68\% and 95\% CQR bands are shown for the total cross section at N$^2$LO order using the EKM potential with $R_0= 0.9\ \text{fm}$. The conformal bands widen gradually with increasing energy, capturing the uncertainty associated with the EFT predictions at higher energy. Looking at Fig. \ref{sgt-cp-intervals}, we can compare the CQR intervals for the total cross section $\sigma_{\text{tot}}$ at N$^2$LO obtained using the pointwise model (left panel) and GP model (right panel), across a range of energies. Both models yield prediction intervals that widen with energy. However, a notable difference is that the GP-based intervals are generally narrower than those from the pointwise model, particularly at higher energies. Additionally, CQR bands based on GP are larger than those of the pointwise model for lower energies. Finally, we note that the bands appear jagged in both cases due to fluctuations in the underlying samples; this effect is stronger for the pointwise model, where samples introduce more noise, and weaker for the GP model, where correlations across energies are taken into account.  

To further explore the versatility of CP, we apply it to the differential cross section $\frac{d\sigma}{d\Omega}$ as a function of polar scattering angle at a fixed energy of $E=96\ \text{MeV}$. For this, we model the differential cross section expansion coefficients as random functions using a GP, allowing us to generate samples of the differential cross sections across the different polar angle ranges. The resulting $68\%$ and $95\%$ CQR intervals, shown in Fig. \ref{dsg-CP-bands}, highlight the ability of CP to provide robust, distribution-free uncertainty quantification for a variety of nuclear observables beyond total cross sections. We emphasize that this approach can be generalized to other observables of interest. 
\begin{figure*}
\centering
	\includegraphics[width=0.95\textwidth]{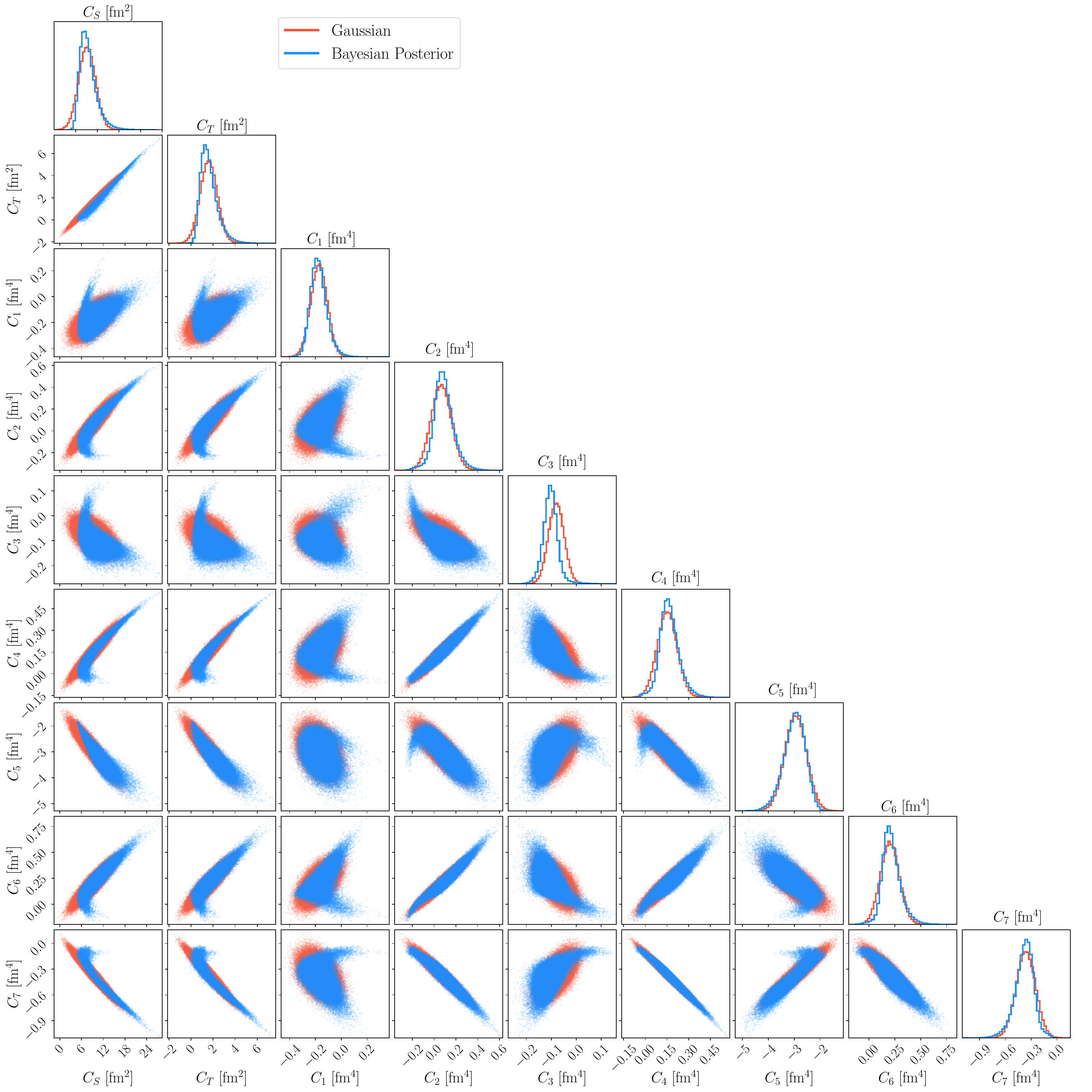} 
	\caption{Corner plot showing the posterior distributions of LECs from Ref.~\cite{Somasundaram_Lynn_Huth_etal_2024} as well as our 
    LECs that have been chosen to have Gaussian distributions with mean values given by the least-squares fit of Ref.~\cite{Somasundaram_Lynn_Huth_etal_2024} and covariance chosen such that they are similar to those coming from the Bayesian framework. We show the distributions for all nine LECs that appear in the local two-body interaction at N$^2$LO, for a coordinate space cutoff of $R_0 = 0.6\ \text{fm}$.}
	\label{cornerplot}
\end{figure*}
\begin{figure*}
\centering
	\includegraphics[width=0.98\textwidth]{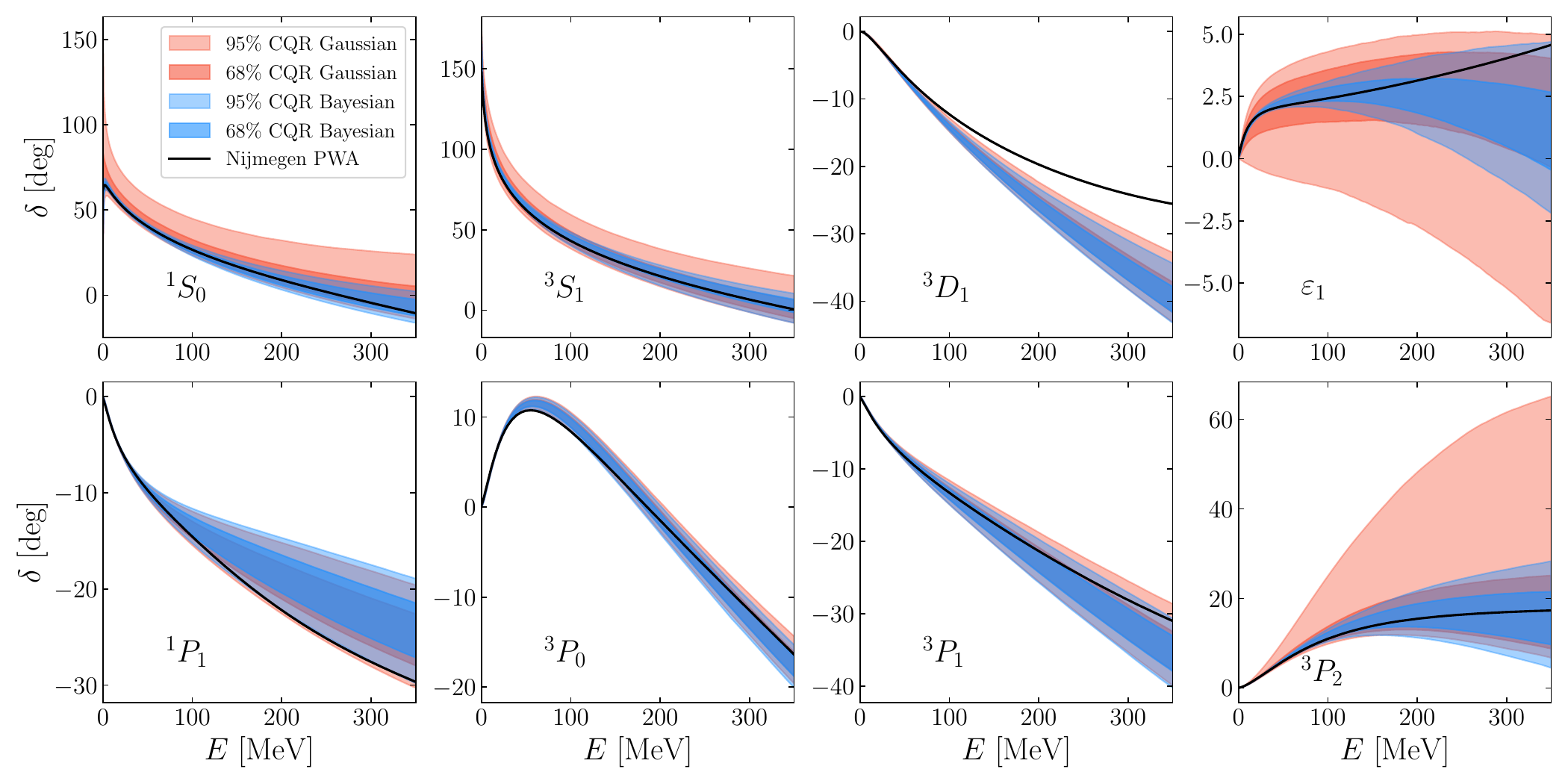} 
	\caption{ 95\% and 68\% CQR intervals for scattering phase shifts in different partial wave channels as a function of energy, generated using the N$^2$LO interactions as described in the main text. The CQR bands are constructed from 2500 samples and adapt to the spread of the data, with the black curve showing the Nijmegen partial-wave analysis \cite{Stoks_Klomp_Rentmeester_etal_1993}. Since the underlying raw phase shift samples are smooth, the resulting CQR bands also appear smooth, in contrast to jagged bands observed for the total cross section.}
	\label{phaseshifts-bands-95,68}
\end{figure*}

\begin{figure*}
\centering
	\includegraphics[width=0.98\textwidth]{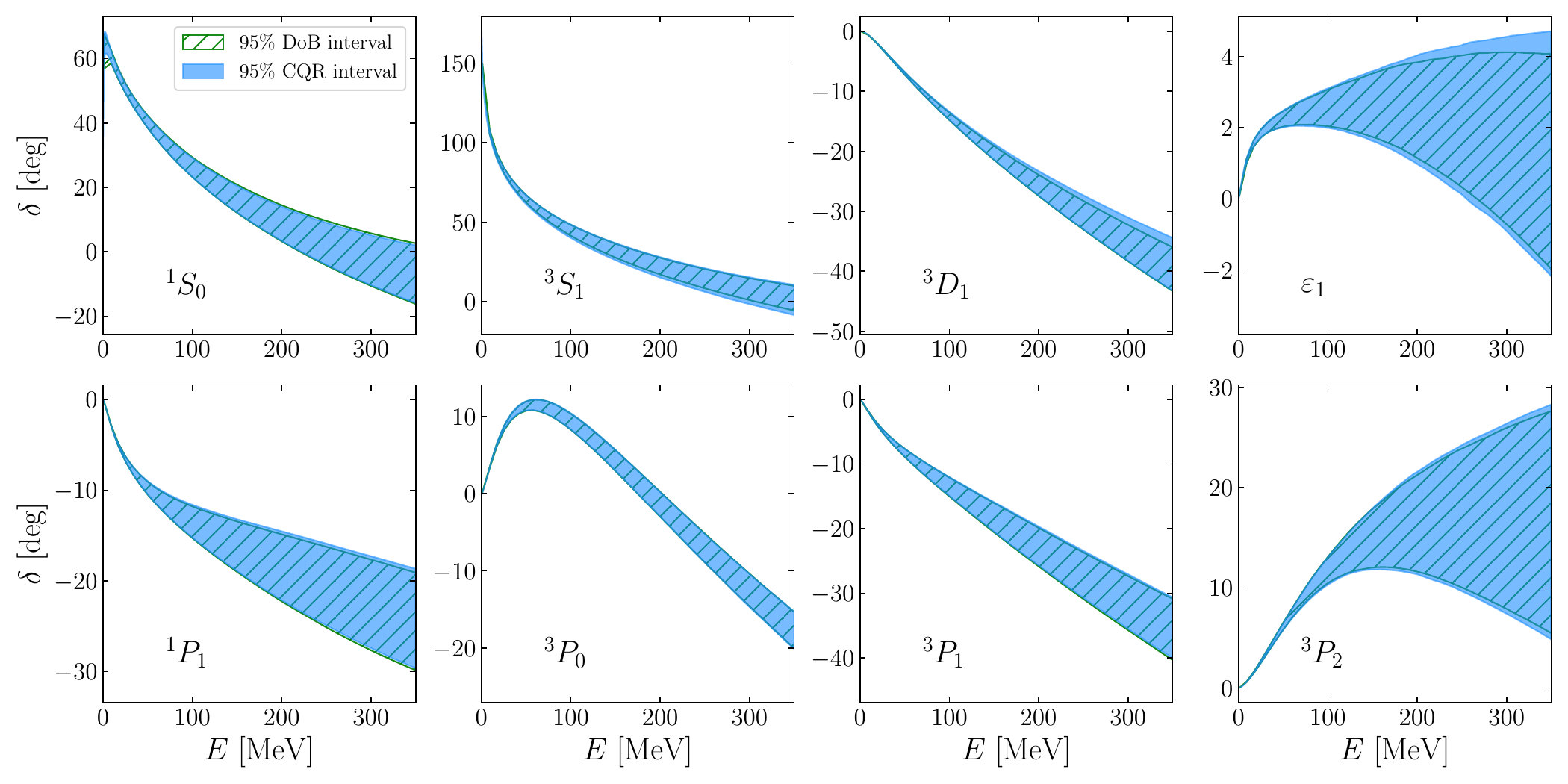} 
	\caption{ Comparison of our 95\% conformal prediction interval (blue shaded region) with the 95\% credible interval (green hatched region) as reported in Ref.~\cite{Somasundaram_Lynn_Huth_etal_2024}. We see very good agreement between our conformal prediction results and the credible interval over all partial wave channels. Interestingly, we do see some slight widening of the uncertainty bands for larger energies in certain channels. }
	\label{CP_vs_Bayes}
\end{figure*}

\subsection{Phase shifts} \label{sec:phaseshifts}
In the previous two sections, we have considered a single interaction from chiral EFT in order to show the utility of our conformal-prediction approach in the analysis of nucleon-nucleon scattering observables. Here, we instead choose to consider an entire family of local next-to-next-to-leading order (N$^2$LO) interactions \cite{Gezerlis_Tews_Epelbaum_etal_2013, Gezerlis_Tews_Epelbaum_etal_2014, Somasundaram_Lynn_Huth_etal_2024} that are used in nuclear many-body methods such as quantum Monte Carlo. In Ref.~\cite{Somasundaram_Lynn_Huth_etal_2024} they report on a Bayesian inference procedure used to fit the nucleon-nucleon interaction and produce posterior distributions for the low-energy couplings (see below). These posterior distributions represent a family of nuclear interactions that all reproduce nucleon-nucleon scattering data. In what follows, we apply our conformal-prediction architecture to nucleon-nucleon phase shifts calculated using this group of interactions. We have also repeated the process for a set of interactions generated \textit{ad hoc}, where we have assumed naive Gaussian distributions for the low-energy couplings in order to explore how our approach handles different types of parameter distributions. The posterior distributions from Ref.~\cite{Somasundaram_Lynn_Huth_etal_2024} are compared with our naive Gaussian distributions in Fig.~\ref{cornerplot}, where it can be seen that we have chosen our LECs such that they have similar covariance to the full Bayesian posteriors.

A realistic description of the nucleon-nucleon interaction is complicated by the fact that it does not depend on only the two-particle separation. It includes terms that depend on the momenta of the nucleons, their spin/isospin states, as well as spin-orbit and tensor contributions. 
At N$^2$LO, the short-range part of the nuclear interaction is written as the sum of contact operators that appear at previous orders,

\begin{align}
V^{\text{LO}} &= C_S \mathbb{1}  + C_T \bm{\sigma}_1 \cdot \bm{\sigma}_2 
\\
V^{{\text{NLO}}} &= C_1  q^2 + C_2 q^2 \bm{\tau}_1 \cdot \bm{\tau}_2 \nonumber
\\
&\ \ \ + C_3 q^2 \bm{\sigma}_1\cdot\bm{\sigma}_2 + C_4q^2 \bm{\sigma}_1 \cdot \bm{\sigma}_2 \bm{\tau}_1 \cdot \bm{\tau}_2 \nonumber
\\
&\ \ \ + \frac{iC_5}{2}(\bm{\sigma}_1 + \bm{\sigma}_2)\cdot \bm{q} \times \bm{k} + C_6 (\bm{\sigma}_1 \cdot \bm{q})(\bm{\sigma}_2\cdot\bm{q}) \nonumber
\\
&\ \ \ + C_7(\bm{\sigma}_1 \cdot \bm{q})(\bm{\sigma}_2\cdot\bm{q}) \bm{\tau}_1\cdot\bm{\tau}_2 , 
\end{align}
where the $C_S$, $C_T$, and $C_{1-7}$ are the low-energy couplings (LECs) that capture the relative contribution of the different operators. The momentum transfer in the exchange channel $\bm{k}$ is an example of the aforementioned momentum dependence and the $\bm{\sigma}$ and $\bm{\tau}$ are spin and isospin operators. 

In order to use these nuclear interactions, the LECs are fit to reproduce experimental two-body scattering data, typically in the form of partial-wave decomposed phase shifts \cite{Stoks_Klomp_Rentmeester_etal_1993, NavarroPerez_Amaro_RuizArriola_2013}. For a given interaction, one can compute the nucleon-nucleon phase shifts by solving the radial Schr\"{o}dinger equation in terms of the two-particle separation,
\begin{align}
    \biggl[-\frac{d^2}{dr^2} + \frac{l(l+1)}{r^2} + \frac{2m}{\hbar^2}V(r) \biggl] u_l(r)  = k^2 u_l(r),
\end{align}
where $k=\sqrt{2mE}/\hbar$, satisfying the boundary conditions that the wavefunction must be zero at the origin and match the asymptotic behavior (i.e. a linear combination of spherical Bessel functions) in the limit of large $r$. A physical intuition for the phase shift $\delta_l$ can be found by comparing the behavior of the scattered particle's wavefunction in the limit of large separation,
\begin{align}
    u_l(r \rightarrow \infty) \propto A_l \sin(kr - \frac{1}{2}l\pi + \delta_l),
\end{align}
and the asymptotic behavior of the partial wave expansion of a plane wave where it can be seen that the phase shift $\delta_l$ captures how the scattering event has modified the behavior of particle's wavefunction. To solve for the nucleon-nucleon phase shifts we have employed the Numerov method across a range of the lab-frame energy in different partial wave channels.  

\begin{figure} [b]
\centering
	\includegraphics[width=0.5\textwidth]{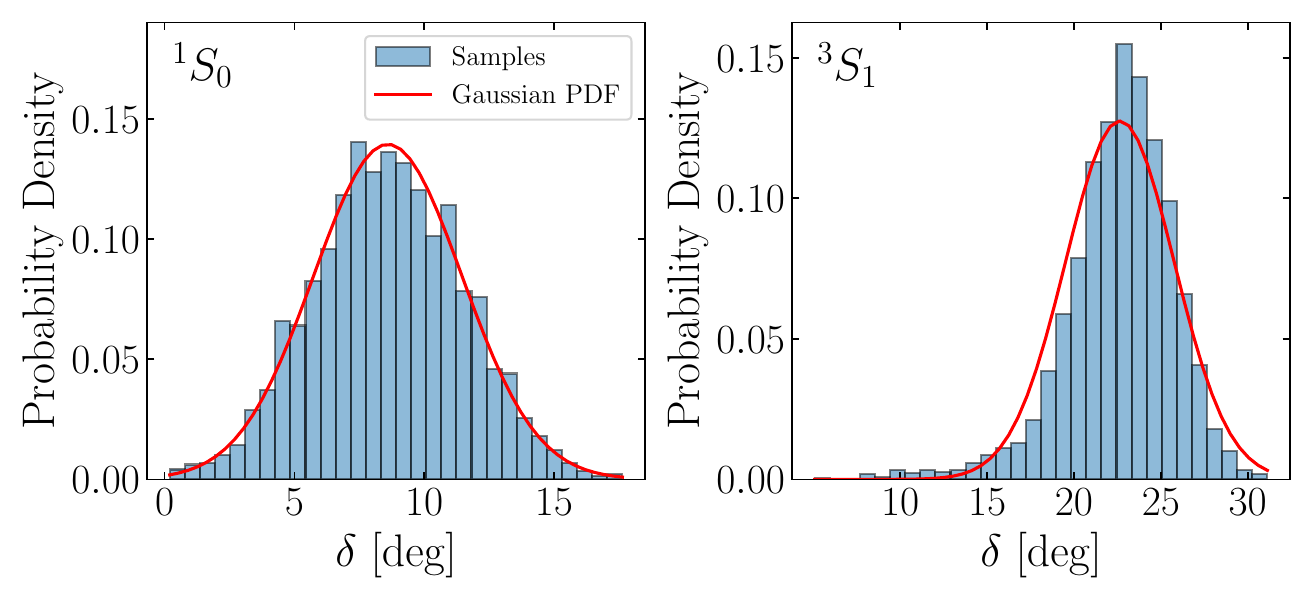} 
	\caption{Histogram of phase shifts calculated using the LECs drawn from the Bayesian posteriors in Ref.~\cite{Somasundaram_Lynn_Huth_etal_2024}, at $E=200\ \text{MeV}$ for the ${}^1S_0$ and ${}^3S_1$ partial waves, compared against Gaussian probability density functions. The ${}^1S_0$ partial wave shows approximately Gaussian behavior, while ${}^3S_1$ has considerable skewness. The empirical coverage for the ${}^1S_0$ and ${}^3S_1$ partial waves is 0.94844 and 0.94870, respectively for the 2500 samples. }
	\label{phase_all_dist}
\end{figure}

In Ref.~\cite{Somasundaram_Lynn_Huth_etal_2024} the LECs at N$^2$LO are fit to reproduce the phase-shift data from the Nijmegen partial-wave analysis (PWA) \cite{Stoks_Klomp_Rentmeester_etal_1993} for laboratory-frame energies up to $150$ MeV in the  $^1\text{S}_0$, $^3\text{S}_1$, $^1\text{P}_1$, $^3\text{P}_0$, $^3\text{P}_1$, and $^3\text{P}_2$ partial-wave channels and the $\epsilon_1$ mixing angle. Therefore, we compute the phase shifts for each of these channels, as well as the coupled $^3$D$_1$ channel, up to $350\ \text{MeV}$ and use the resulting distribution of phase shifts in order to construct conformal-prediction uncertainty bands. 

In Fig. \ref{phaseshifts-bands-95,68} we show both 95\% and 68\% CQR intervals for the phase shifts across a range of energies and different partial waves. These prediction bands are constructed based on samples and reflect the uncertainty in the EFT predictions. Because conformal prediction is distribution-free, the bands remain valid even in the presence of skewed or non-Gaussian distributions, making them rigorous uncertainty quantification tools in nucleon-nucleon scattering. We can see that by using our conformal-prediction approach we are able to construct uncertainty bands that capture the data regardless of the underlying distribution and maintain the desired level of empirical coverage. It is interesting to note that the CQR bands for the $p$-wave channels seem to be largely unchanged between the two sets of interactions considered, and that the CQR bands in the $s$-wave channels show a clear dependence, with considerable broadening for the less-physical family of interactions. Finally, it is worth noting that the phase shift intervals show a smooth behavior, which directly reflects the smoothness of the underlying samples. This contrasts with the total cross section case, where noisier samples lead to jagged intervals.

In Fig.~\ref{CP_vs_Bayes} we also compare the uncertainty bands generated by our conformal prediction approach against the credible interval results reported in Ref.~\cite{Somasundaram_Lynn_Huth_etal_2024}. While the conformal prediction approach is indeed a postprocessing technique, in this case applied to the phase shifts calculated using the posterior distributions from Ref.~\cite{Somasundaram_Lynn_Huth_etal_2024}, it is very reassuring to see that we produce uncertainty bands that agree with these previous results. We do see some broadening in the larger energies for certain partial waves, which is indicative of how the uncertainty bands change with guaranteed coverage. It would be interesting in future work to attempt a wedding of these two techniques, where perhaps the constraint of imposing a set level of empirical coverage could be imposed in the Bayesian framework. 
\begin{figure} 
\centering
	\includegraphics[width=0.5\textwidth]{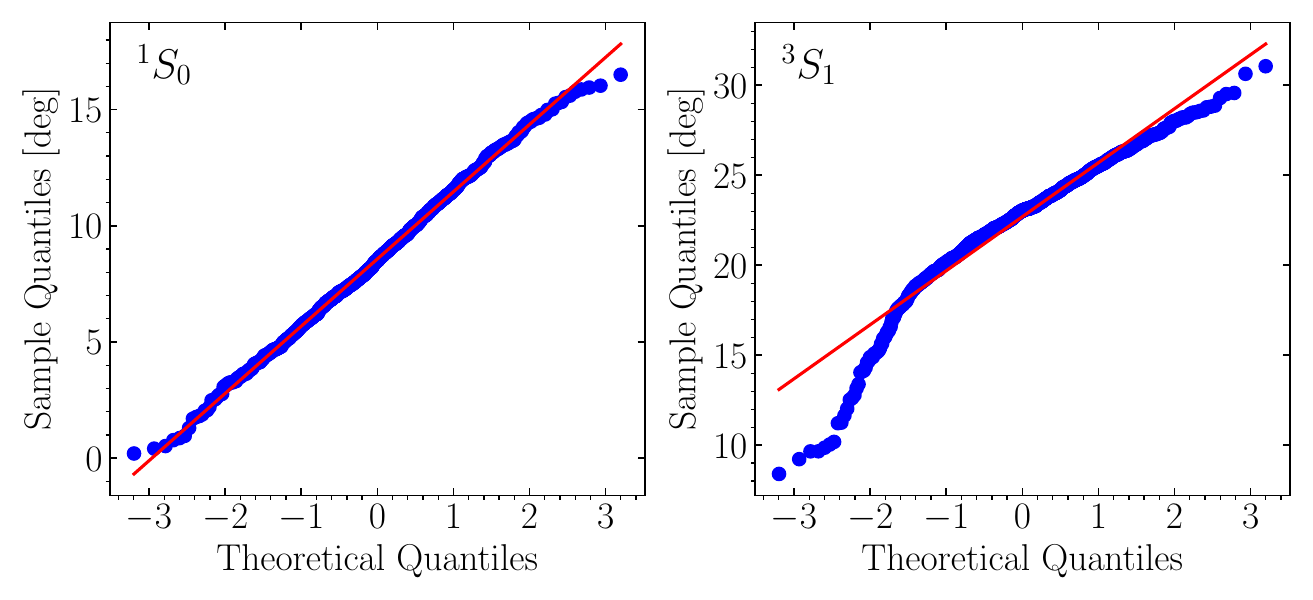} 
	\caption{Q-Q plot comparing the empirical quantiles of 2500 phase shift samples for the ${}^1S_0$ and ${}^3S_1$ partial waves computed using LECs drawn from the Bayesian posterior of Ref.~\cite{Somasundaram_Lynn_Huth_etal_2024}, to the theoretical quantiles of a normal distribution. The deviations from the red diagonal line in the ${}^3S_1$ highlights the non-Gaussian behavior of the data, supporting the need for distribution-free methods like conformal prediction. }
	\label{QQ}
\end{figure}
To further investigate the statistical characteristics of the chiral EFT phase-shift predictions, we analyze the behavior of phase-shift samples found using the Bayesian posterior LECs at E $=200\ \text{MeV}$, for the ${}^1S_0$ and ${}^3S_1$ partial waves. We visualize their distributions using histograms and compare with Gaussian probability density functions (see Fig. \ref{phase_all_dist}). While the ${}^1S_0$ partial wave appears to have a distribution close to Gaussian, the ${}^3S_1$ shows skewness or heavy tails. We find similar skewed behavior for both the $^3D_1$ and $^3P_2$ partial waves, indicating that this behavior could be due to the coupled nature of these partial wave channels. 

This is further confirmed by the quantile-quantile (Q-Q) plot, as shown in Fig.~\ref{QQ}. A Q-Q plot is a diagnostic tool used to assess whether a dataset follows a specific theoretical distribution, in this case, the standard normal distribution. It compares the empirical quantiles from a reference distribution. In the Q-Q plot shown in Fig.~\ref{QQ}, the $x$ axis represents the theoretical quantiles of a standard normal distribution, these are calculated using
\begin{align}
    x_i = \Phi^{-1}\biggl(\frac{i -0.5}{n}\biggl),
\end{align}
where $\Phi^{-1}$ denotes the quantile function of the standard normal distribution, $n$ is the sample size, and $i$ is the rank of the ordered samples. These values are dimensionless and represent the expected quantiles if the data were normally distributed. The y-axis shows the empirical quantiles of the data, which are simply the ordered phase shift samples at a fixed energy. These values retain their physical units (degrees), as they correspond directly to the sampled phase shift values. The red line in each plot serves as a reference for perfect agreement with the normal distribution.  If the data are normally distributed, the plotted points lie approximately along the $45$ degree line. Deviations from this line indicate departures from normality.  As we can see in Fig.~\ref{QQ}, for the ${}^3S_1$ partial wave, the quantiles deviate significantly from the straight line, particularly in the tails. This deviation shows the presence of skewness in the data. In spite of this skewness, the empirical coverage for the 95\% prediction interval at $E = 200\ \text{MeV}$ is essentially identical for both the $^1S_0$ and $^3S_1$ partial waves (0.94844 and 0.94870 respectively).
It is important to reiterate that the CP method does not rely on any assumptions about the underlying distribution. CP constructs prediction intervals regardless of whether the data follow a Gaussian or any other distribution. 

Our approach of generating CP uncertainty bands for nucleon-nucleon phase shifts could in principle be used as a new constraint when fitting interaction parameters to reproduce experimental data. This also highlights one of the key strengths of the CP approach, in that this process of enforcing guaranteed coverage is agnostic as to whether the fitting procedure is carried out using a frequentist or Bayesian framework, which should prove very useful in future work on the nucleon-nucleon force.

\section{Summary and Outlook}
In this work, we developed a distribution-free approach for uncertainty quantification in the nucleon-nucleon scattering problem for physical observables, such as phase shifts and cross sections. Our method uses CP as a postprocessing step applied to samples drawn from the Bayesian posteriors. CP provides finite-sample prediction intervals with guaranteed coverage, regardless of the underlying data distribution. We demonstrated this capability by applying CP to three different scenarios; the pointwise model, the GP model, and phase shifts calculated for local interactions with normally distributed LECs. In the pointwise model, we assumed that the expansion coefficients in the EFT series are sampled from a Student-t distribution, and generated samples for total cross sections at N$^2$LO for four different energies, then we showed the 95\% and 68\% CQR bands for the observable. We also calculated the empirical coverage for a range of $1-\alpha$, and computed that our method guarantees the desired coverage. Then, using a Gaussian process, we modeled the expansion coefficients as random functions. The resulting samples at N$^2$LO demonstrated approximately Gaussian behavior, and CQR was applied to construct 95\% prediction intervals across different ranges of energies. And finally, in the phase-shift analysis, we explored the behavior of partial- wave samples across different laboratory energies. Histograms and Q-Q plots revealed that while the ${}^1S_0$ partial wave followed a near- Gaussian distribution, the $^3S_1$ shows skewness. Despite these deviations from normality, CQR intervals remained valid. By presenting both 68\% and 95\% CQR intervals, we showed how the method can quantify uncertainties. Overall, our results highlight the strength of conformal prediction as a powerful tool for uncertainty quantification. By combining CP with posterior samples, we have established a robust framework with minimal assumptions. We have also checked for both scattering data and nucleon-nucleon phase shifts that our conformal prediction approach provides uncertainty bands that agree well with previously published Bayesian analyses. We hope in the future to build upon this work and develop a framework which marries these two approaches. 

Beyond nucleon-nucleon scattering, CP also shows promise in astrophysics. Recent work has demonstrated the potential of CP, where it has been used to calibrate search algorithms for gravitational waves \cite{Ashton_Colombo_Harry_etal_2024}. Looking ahead, one could use CP to quantify uncertainty in the EOS of the neutron stars, or as a new constraint in the fitting of nucleon-nucleon interactions for use in many-body calculations. In addition, it could be possible to include CP uncertainty bands into existing frameworks that use emulators for uncertainty quantification in many-body observables \cite{Wesolowski_Svensson_Ekstrom_etal_2021, Jiang_Forssen_Djarv_etal_2024a, Somasundaram_Armstrong_Giuliani_etal_2025, Armstrong_Giuliani_Godbey_etal_2025, Curry_Hebeler_Gandolfi_etal_2025}. While existing work employs Bayesian inference, an approach utilizing CP would offer a coverage guarantee under minimal assumptions, particularly valuable in the face of limited or model-dependent data.

\section*{Acknowledgments}
The authors would like to thank Ingo Tews and Rahul Somasundaram for sharing their LEC posterior distributions and credible interval bands from Ref.~\cite{Somasundaram_Lynn_Huth_etal_2024}, as well as the BUQEYE collaboration for making their source code publicly available. We would also like to thank ECT* for support at the workshop \textit{Next generation ab initio nuclear theory} during which this work was discussed.
This work was supported by the Natural Sciences and Engineering Research Council (NSERC) of Canada and the Canada Foundation for Innovation (CFI). Computational resources have been provided by Compute Ontario through the Digital Research Alliance of Canada, and by the National Energy Research Scientific Computing Center (NERSC), which is supported by the U.S. Department of Energy, Office of Science, under Contract No. DE-AC02-05CH11231.

\bibliography{bib}
\end{document}